\documentclass[12pt]{article}

\usepackage{amsfonts,amssymb,graphicx,cite}

\newcommand{\appsection}{\addtocounter{section}{1}\setcounter{equation}{0}
}

\newcommand{\ie}{{\it i.e.}}

\newcommand{\eg}{{\it e.g.}}

\newcommand{\etc}{{\it etc.}}
\newcommand{\eq}{Eq.}
\newcommand{\eqs}{Eqs.}
\newcommand{\fig}{Fig.}
\newcommand{\figs}{Figs.}
\newcommand{\Ref}{Ref.\,}
\newcommand{\Refs}{Refs.\,}
\newcommand{\Sec}{Sec.}

\newcommand{\be}{\begin{equation}}
\newcommand{\ee}{\end{equation}}

\begin{document}

\bibliographystyle{elsart-num}

\topmargin 0pt
\oddsidemargin=-0.4truecm
\evensidemargin=-0.4truecm
\renewcommand{\thefootnote}{\fnsymbol{footnote}}

\newpage

\setcounter{page}{1}

\begin{titlepage}     
\vspace*{-2.0cm}  
\begin{flushright}
TUM-HEP-413/01\\
MPI-PhT/01-11\\
hep-ph/0105029
\end{flushright}
\vspace*{0.5cm}
\begin{center}
\vspace*{0.2cm}

{\Large \bf T violation in neutrino oscillations in matter}
\\
\vspace{1.0cm}
{\large Evgeny Akhmedov
\footnote{On leave from National Research Centre Kurchatov Institute, 
Moscow 123182, Russia. E-mail address: {\tt akhmedov@physik.tu-muenchen.de}}, 
Patrick Huber
\footnote{E-mail address: {\tt phuber@physik.tu-muenchen.de};
Max-Planck-Institut f\"ur Physik, Postfach 401212, D-80805 M\"unchen,
Germany},
Manfred Lindner
\footnote{E-mail address: {\tt lindner@physik.tu-muenchen.de}}, \\
Tommy Ohlsson
\footnote{E-mail addresses: {\tt tohlsson@physik.tu-muenchen.de}, 
{\tt tommy@theophys.kth.se}; Division of Mathematical Physics,
Theoretical Physics, Department of Physics, Royal Institute of
Technology, SE-100 44 Stockholm, Sweden}}

\vspace{0.1cm}
{\em Institut f{\"u}r Theoretische Physik, Physik Department,
Technische Universit{\"a}t M{\"u}nchen James-Franck-Stra{\ss}e,
D-85748 Garching bei M{\"u}nchen, Germany}

\end{center}

\vglue 1.2truecm

\begin{abstract} 
We consider the interplay of fundamental and matter-induced T violation 
effects in neutrino oscillations in matter. After discussing the general 
features of these effects we derive a simple approximate analytic expression 
for the T-violating probability asymmetry $\Delta P^T_{ab}$ for three-flavour 
neutrino oscillations in a matter with an arbitrary density profile in terms 
of the two-flavour neutrino amplitudes. Explicit examples are given for the 
cases of a two-layer medium and for the adiabatic limit in the general case. 
We then discuss implications of the obtained results for long baseline 
experiments. We show, in particular, that asymmetric matter effects cannot 
hinder the determination of the fundamental CP and T-violating phase 
$\delta_{CP}$ in the long baseline experiments as far as the error in this 
determination is larger than 1\% at 99\% C.L. Since there are no
T-violating effects in the two-flavour case, and in the limits of vanishing 
$\theta_{13}$ or $\Delta m_{21}^2$ the three-flavour neutrino oscillations 
effectively reduce to the two-flavour ones, studying the T-violating 
asymmetries $\Delta P^T_{ab}$ can in principle provide us with a 
complementary means of measuring $\theta_{13}$ and $\Delta m_{21}^2$. 
\end{abstract}
\vspace{1.cm}
\centerline{{\em PACS:} 14.60.+Pq; 26.65.+t} 
\vspace{.3cm}
\centerline{{\em Keywords:} Neutrino oscillations; matter effects;
T violation} 
\end{titlepage}

\renewcommand{\thefootnote}{\arabic{footnote}}
\setcounter{footnote}{0}
\newpage

\section{Introduction}

The experimental evidence for neutrino oscillations implies that 
neutrinos are massive and that mixing angles and CP phases exist
in the lepton sector. The measurement of these neutrino parameters 
is very important, not only because it will provide us with an information
about neutrino properties, but also because it may have interesting 
implications for the structure of theories at very high energies, which  
might ``explain'' the low energy parameters. Since neutrinos could in 
general have both Dirac and Majorana masses, they might, in principle,
also allow more insight into the flavour problem than quarks do.
Extraction strategies for neutrino masses and mixings in current 
and future experiments are therefore an important subject under 
study. Future experiments also offer to study CP, T, and CPT  
properties in the neutrino sector. Local quantum field theories 
are, however, for general reasons invariant under CPT, and any 
CP violation implies then a correlated T violation. An independent 
study of T violation might therefore not appear very interesting, 
unless it is sensitive to tiny CPT violating effects induced at 
very high energy scales by physics going beyond local quantum field 
theory. In this paper, we discuss that this is different for 
neutrinos and that an independent study of T violation for neutrino 
oscillations in matter offers interesting insights. The point is 
that the presence of matter in the experimental setup violates  
by itself CP as well as CPT and gives thus rise to an extra 
CP violation in addition to the intrinsic CP violation. Furthermore, 
there are extra T-violating effects if the matter density profile seen by 
neutrinos is asymmetric. This extra T violation does not follow directly
from CPT because CPT itself is violated by the presence of matter.

One can understand these effects by comparing CP, T, and CPT 
properties of neutrino oscillations in vacuum and in matter. 
Using CPT one finds in vacuum 
$P(\nu_a\rightarrow\nu_b) = P(\bar\nu_b\rightarrow\bar\nu_a)$.
This is no longer true when one studies neutrino oscillations in 
matter, \ie, a CP- and CPT-asymmetric environment, 
where the oscillation probabilities of neutrinos and antineutrinos 
change in a different way due to the differences in coherent 
forward scattering in a given medium (MSW effect).  
As a consequence, CPT is violated by matter effects, \ie, 
$P(\nu_a\rightarrow\nu_b) \neq P(\bar\nu_b\rightarrow\bar\nu_a)$.
Moreover, the total CP violation is now a combined effect, where intrinsic 
CP violation must be separated in analyses from effects induced by 
the CP-violating environment. A quantity which measures the total
CP violation is $\Delta P^{CP}_{ab} = P(\nu_a\rightarrow\nu_b) - 
P(\bar\nu_a\rightarrow\bar\nu_b)$.
Such ``CP differences'' or corresponding asymmetries have been studied for 
appearance channels like $\nu_e\rightarrow \nu_\mu$ at future long baseline 
experiments (see, \eg, \Refs \cite{Cervera,Freund:2000ti,Barger01}), and
one can nicely see  how the CP difference is in general a combination of 
intrinsic CP effects and matter effects, while in vacuum $\Delta P^{CP}_{ab}$ 
depends only on intrinsic CP violation. Note that one can easily see from 
CPT and the definition of $\Delta P^{CP}_{ab}$ that in vacuum CP violation 
can only occur in appearance channels, \ie, for $a \neq b$, while in matter 
one has in general $\Delta P^{CP}_{aa}\neq 0$. 

In analogy to the CP difference $\Delta P^{CP}_{ab}$ one can 
also define a ``T difference'' $\Delta P^{T}_{ab} = 
P(\nu_a\rightarrow \nu_b) - P(\nu_b\rightarrow\nu_a) $
and the ``CPT difference'' $\Delta P^{CPT}_{ab} = 
P(\nu_a\rightarrow\nu_b) - P(\bar\nu_b\rightarrow\bar\nu_a)$.
In vacuum, where CPT holds, one has $\Delta P^{CP}_{ab} + 
\Delta P^T_{\bar a\bar b} = \Delta P^{CPT}_{ab} \equiv 0$
and $\Delta P^T_{\bar a\bar b}$ is thus given directly by 
$\Delta P^{CP}_{ab}$. 
Since in matter CPT is no longer valid, one now has  
$\Delta P^{CPT}_{ab}\neq 0$. 
For this reason, T violation is no longer correlated with CP violation. 
In particular, matter does not necessarily break T invariance; as was
mentioned above, this only happens if its density profile seen by 
neutrinos is asymmetric. 

T violation in neutrino oscillations in matter has been studied in a number 
of papers (see, \eg, \Refs \cite{KP,KrPe,T1,Arafune,Koike97,Bil98,Barger99,
Koike00,Yasuda99,Sato00,Koike00a,HS,Andre,YKT,PW,Koike00b,MTKY}).
However, in most of these studies matter of constant density was considered, 
where only intrinsic T violation is possible. T violation in neutrino 
oscillations in non-uniform matter was considered in \Refs \cite{Andre,MTKY}.  
In the first of these papers T-violating effects of solar matter on the
oscillations of ${\cal O}$(GeV) neutrinos, which could be produced in the 
annihilation of WIMPs inside the sun, are considered, whereas in the 
second paper small perturbations of constant matter density profiles 
were discussed. In the present paper we study T violation in neutrino
oscillations in matter of an arbitrary density profile and discuss the 
interplay of the fundamental and matter-induced T violation. In particular, 
we discuss where the matter-induced T violation effects may play a role
and where they can be safely ignored. 

The paper is organized as follows. In \Sec~\ref{sec:general} 
we discuss the general features of T violation in neutrino 
oscillations in matter. Subsequently, we give approximate 
analytic results for arbitrary matter density profiles in 
\Sec~\ref{sec:approx} and discuss the implications for
long baseline experiments, solar, atmospheric, supernova, and
cosmological neutrinos in \Sec~\ref{sec:impl}. We discuss the 
obtained results and conclude in \Sec~\ref{sec:dc}. 
In Appendix A we give details of our general analytic approach, whereas in
Appendix B its application to two particular cases is considered.

\section{General features of T violation in neutrino oscillations in
matter}
\label{sec:general}

Oscillations of neutrino flavour in vacuum or in matter are described by
the Schr\"odinger equation
\be
i\frac{d}{dt}|\nu\rangle = H(t)|\nu\rangle\,,
\label{Sch1}
\ee
where $|\nu\rangle$ is the neutrino vector of state and $H(t)$ is the 
effective Hamiltonian which in general depends on time $t$ through the 
$t$-dependence
\footnote{We consider the evolution of the neutrino system in time rather 
than in space (since for relativistic neutrinos $r\simeq t$, both
descriptions are equivalent). The coordinate dependence of the
matter density profile $N(r)$ therefore translates into its time
dependence $N(t)$.} 
of the matter density $N(t)$: $H(t) \equiv H[N(t)]$
\footnote{Neutrino flavour evolution in matter depends on the total matter
density as well as on the electron number fraction $Y_e$. Throughout this 
paper we will for brevity call them collectively ``matter density profile''.}.

We will assume that neutrinos are stable and are not absorbed in matter;
in this case the Hamiltonian $H(t)$ is Hermitian. 
It is convenient to define the evolution matrix $S(t,t_0)$:
\be
|\nu(t)\rangle=S(t,t_0)|\nu(t_0)\rangle\,.
\label{S1}
\ee 
It has the obvious properties
\be
S(t,t_0)=S(t,t_1)S(t_1,t_0)\,,\qquad
S(t_0,t_0)={\bf 1}\,,\qquad 
S(t,t_0)S(t,t_0)^\dagger={\bf 1}\,,
\label{S2}
\ee
where the last property (unitarity) follows from the hermiticity of $H(t)$.  
{}From \eqs~(\ref{Sch1}) and (\ref{S1}) it follows that $S(t,t_0)$ satisfies
the equation 
\be
i\frac{d}{dt}S(t,t_0) = H(t)S(t,t_0)\,.
\label{Sch2}
\ee
It is also sometimes useful to consider the evolution equation of $S(t,t_0)$ 
with respect to its second argument, $t_0$. 
Differentiating the equality $S(t,t_0)S(t_0,t)={\bf 1}$ with respect to 
$t_0$ and using \eq~(\ref{Sch2}), one finds 
\be
i\frac{d}{dt_0}S(t,t_0) = -S(t,t_0)H(t_0)\,.
\label{Sch2a}
\ee
The amplitudes of neutrino flavour transitions are just the 
elements of the evolution matrix $S$:
\be
A[\nu_a(t_0)\to\nu_b(t)]=[S(t,t_0)]_{ba}\,.
\label{S3}
\ee 

We are interested in the properties of the solutions of the evolution 
equations (\ref{Sch1}) or (\ref{Sch2}) with respect to the time reversal 
transformation T. Unlike in the case of CP transformation, these properties 
cannot be directly experimentally tested as one cannot 
change the direction of time. However, time reversal can be 
studied by simply interchanging the initial and final neutrino flavours.
This can be readily seen in the case of neutrino oscillations in vacuum, for 
which
\be
[S(t,t_0)]_{ba}=\sum_j U_{aj} \,e^{-iE_j(t-t_0)}\,U_{bj}^*\,,
\label{Svac} 
\ee
where $U$ and $E_i$ are the lepton mixing matrix and energy eigenvalues in
vacuum, respectively. Indeed, time reversal interchanges the initial
and final times, $t_0$ and $t$, respectively, which leads to the
complex conjugation of the exponential factors in \eq~(\ref{Svac}). On
the other hand, the interchange of the initial and final neutrino 
flavours $a\leftrightarrow b$ (\ie, $S\to S^T$, where the superscript 
$T$ denotes transposition) is equivalent to the complex conjugation of the 
matrix elements of the leptonic mixing matrix $U$ in \eq~(\ref{Svac}). The 
action of these two operations on the elements of the evolution matrix $S$ in 
\eq~(\ref{Svac}) differ from each other only by complex conjugation. Since the 
transition probabilities $P_{ab}\equiv P(\nu_a\to\nu_b)$ are just the squares 
of the moduli of the matrix elements $S_{ba}$, these two procedures are 
physically equivalent. This means, as is well known, that in vacuum, instead 
of studying neutrino oscillations ``backward in time'', one can study the
oscillations forward in time, but with the initial and final 
flavours interchanged. 

The situation is, however, less obvious in the case of neutrino oscillations 
in matter of non-constant (and, in general, asymmetric) matter density. Time
reversal means, in particular, that the matter density profile has to be 
traversed in the opposite direction. That is to say, one has to consider the 
evolution in the ``reverse'' profile $\tilde{N}(t)$ 
rather than in the ``direct'' profile $N(t)$. Here $\tilde{N}(t)$ should be
understood as the profile seen by neutrinos when the positions of the 
neutrino source and detector are interchanged.
Therefore, the following question arises: When one reduces the probability
of the time-reversed neutrino oscillations in matter to that for the
interchanged neutrino flavours, does one have to simultaneously replace the 
direct profile by the reverse one? There is some confusion in the literature
regarding this issue, and therefore we believe that it is worth clarifying
it here. In general, in the case of neutrino oscillations in matter of 
non-constant density, no closed-form expressions for the transition
probabilities exist, and the arguments similar to those applied to
\eq~(\ref{Svac}) cannot be used.
There is, however, a very simple and general argument which 
does not depend on whether neutrinos oscillate in vacuum  or in matter and
what the matter density profile is.  

Indeed, under time reversal the arguments of the evolution matrix $S(t,t_0)$ 
are interchanged:
$$
T: \qquad S(t,t_0) \to S(t_0,t)\,.
$$
{}From \eq~(\ref{S2}) one then finds 
\be
S(t_0,t)=S(t,t_0)^{-1}=S(t,t_0)^\dag=[S(t,t_0)^T]^{*}\,.
\label{S5}
\ee
Therefore, the operations of time reversal and interchange of initial 
and final neutrino flavours [\ie, transposition of $S(t,t_0)$] are
related by complex conjugation and so lead to the same transition
probabilities in matter with an arbitrary density profile as
well as in vacuum. This, in particular,
means that instead of considering neutrino oscillations in matter ``backward 
in time'', one can consider the oscillations between interchanged initial
and final neutrino flavours forward in time and in the same (\ie, direct) 
matter density profile. 

In vacuum, due to CPT invariance, T violation is equivalent to CP violation. 
In particular, the CP-odd and T-odd differences of neutrino
oscillation probabilities
$$
\Delta P_{ab}^{CP}\equiv P_{ab}-P_{\bar{a}\bar{b}}\qquad \mbox{and}
\qquad \Delta P_{ab}^{T}\equiv P_{ab}-P_{ba} 
$$
are equal to each other. In the case of $n\ge 3$ neutrino species they are
expressed through $(n-1)(n-2)/2$ Dirac-type CP-violating phases 
$\{\delta_{CP}\}$ 
\footnote{If neutrinos are Majorana particles, there are $n-1$
additional (so-called Majorana) phases, but they have no effect on
neutrino oscillations.}.

The situation with neutrino oscillations in matter is drastically different. 
Ordinary matter is both CP and CPT-asymmetric as it consists of particles
(electrons and nucleons) and not of their antiparticles or, in general, of  
unequal numbers of particles and antiparticles. The violation of CPT by
matter implies that CP and T violation effects are in general different.
CP is violated by the very existence of matter. This violation manifests 
itself even in the two-flavour neutrino systems (or, in general, even in the 
absence of fundamental CP violation, \ie, when $\delta_{CP}=0$) -- matter 
may enhance the oscillations between neutrinos and suppress those between 
antineutrinos, or vice versa \cite{W,MS}. Moreover, the survival pro-
\newpage\noindent
babilities $P_{aa}$, which are CP-symmetric in vacuum due to CPT 
invariance, are no longer CP-symmetric in matter. 

In contrast to this, the presence of matter does not necessarily break  
T invariance. In particular, matter of constant density (or, in general, 
matter with a density profile that is symmetric with respect to the
interchange of the positions of the neutrino source and detector) does not
induce any T violation. In addition, the survival probabilities $P_{aa}$ are 
always T-symmetric, since the initial and final neutrino flavours coincide. 

The point that symmetric matter density profiles do not induce any
T violation seems intuitively rather obvious; here we will give an
explicit proof of this statement and also will derive several
useful properties of the evolution matrix $S(t,t_0)$. 
To study these properties, it is 
convenient to consider the evolution over the symmetric time interval 
$[-t,\,t]$. This does not lead to any loss of generality as any time 
interval can be reduced to a symmetric one by the proper choice of the 
point $t=0$. Since both arguments of the evolution matrix $S(t, -t)$ 
depend on $t$, it does not satisfy \eq~(\ref{Sch2}). It is, however, not
difficult to derive the evolution equation for $S(t, -t)$. Using 
\eq~(\ref{S2}) one can write $S(t, -t)=S(t, 0)S(0, -t)$. The matrix 
$S(t, 0)$ satisfies the usual evolution equation 
\be
i\frac{d}{dt}S(t,0) = H(t)S(t,0)\,.
\label{Sch3}
\ee
Taking its Hermitian conjugate and substituting $-t$ for $t$, one finds 
\be
i\frac{d}{dt}S(0,-t) = S(0,-t)H(-t)\,.
\label{Sch4}
\ee
Using Eqs. (\ref{Sch3}) and (\ref{Sch4}), one finally obtains 
\be
i\frac{d}{dt}S(t,-t) = H(t)S(t,-t)+S(t,-t)H(-t)\,.
\label{Sch5}
\ee
The action of time reversal on this equation is given by the
substitution $t\to -t$. 

Let us now assume that the fundamental CP and T violation is absent, \ie, 
all $\{\delta_{CP}\}=0$. In this case the Hamiltonian of the neutrino
system is real (or can be made real by a rephasing of the neutrino states). 
Since it is real and Hermitian, it is also symmetric: $H^T=H$. Transposition 
of \eq~(\ref{Sch5}) yields
\begin{eqnarray}
i\frac{d}{dt}S(t,-t)^T &=& S(t,-t)^T H(t)^T+H(-t)^T S(t,-t)^T \nonumber \\
&=& S(t,-t)^T H(t)+H(-t) S(t,-t)^T\,.
\label{Sch6}
\end{eqnarray}

The evolution matrix for neutrinos passing through the reverse profile 
$\tilde{N}(t)=N(-t)$ can be obtained from \eq~(\ref{Sch5}) by replacing 
$H(t)$ with $H(-t)$.  
Comparing the resulting equation with \eq~(\ref{Sch6}), one obtains
\be
S_{\rm dir}(t,t_0)^T = S_{\rm rev}(t,t_0) \qquad
(\{\delta_{CP}\}=0)\,.
\label{S7}
\ee
Here the subscripts {dir} and {rev} denote propagation in the 
direct and reverse profiles, $N(t)$ and $\tilde{N}(t)$, respectively, and
we have reinstated the time interval $[t_0,t]$ in the 
arguments of the evolution matrices. From \eq~(\ref{S7}) it immediately
follows that in the case of a symmetric matter density profile the evolution 
matrix $S(t,t_0)$ is symmetric and there is no matter-induced T violation. 
For the particular case of matter consisting of a number of constant density 
layers, \eq~(\ref{S7}) was derived by Fishbane and Kaus \cite{FK}. 

It is easy to generalize \eq~(\ref{S7}) to the case when $\{\delta_{CP}\}
\ne 0$. Following the lines that led to \eq~(\ref{S7}), one obtains, in this 
case,
\be
S_{\rm dir}(t,t_0)^T = \left. S_{\rm rev}(t,t_0) \right|_{\{\delta_{CP}\}\to \,
-\{\delta_{CP}\}} 
\,.
\label{S8}
\ee
This expression generalizes \eq~(13) of \Ref \cite{FK}, which was obtained
for matter consisting of constant-density layers, to the arbitrary matter 
density profile. \eq~(\ref{S8}) has a rather obvious physical meaning. 
It just reflects the fact that  
there are two kinds of effects that contribute to T violation (\ie, to 
the difference between $S$ and $S^T$) in matter -- intrinsic T violation,
due to the non-vanishing CP and T violating phases $\{\delta_{CP}\}$, and 
extrinsic T violation, due to the asymmetry of the density profile with
respect to the interchange of the positions of the neutrino source and
detector. 

\eq~(\ref{S7}) means that in the case of $\{\delta_{CP}\}=0$ the difference 
\be
\Delta P_{ab}\equiv P_{\rm dir}(\nu_a\to\nu_b) - P_{\rm rev}(\nu_b\to\nu_a) 
\label{P1}
\ee
vanishes. Any deviation of this  difference from zero is therefore a measure 
of non-vanishing fundamental CP and T violation and can, in principle, be used 
 for their experimental searches. Fishbane and Kaus \cite{FK} have stressed 
that one can, in principle, probe the effects of $\{\delta_{CP}\}\ne 0$ even 
by studying the survival probabilities $P_{aa}$ ($a\ne e$) if one compares 
these probabilities for direct and reverse profiles. This is a very 
interesting observation, even though the experiments with interchanged 
positions of neutrino source and detector would certainly be difficult to
perform.   

The point that the survival probabilities can be used for looking for 
fundamental CP and T violation is easy to understand. The probabilities 
$P(\nu_a\to\nu_a)$ are T-symmetric as the initial and final neutrino flavours 
coincide. Therefore, the contributions to their T asymmetry coming from the 
fundamental T violation and from the asymmetry of the matter density
profile must cancel each other exactly. This means that by measuring
the asymmetry $P_{{\rm dir}}(\nu_a \to \nu_a) - P_{{\rm rev}}(\nu_a
\to \nu_a)$ (with $a\ne e$) one directly measures, up to the sign, the
asymmetry due to the fundamental CP and T-violating phases $\{\delta_{CP}\}$.

The asymmetry of the $\nu_e$ survival probability $P_{ee}$ with respect to
density profile reversal vanishes (up to the tiny terms due to radiative 
corrections to matter-induced potentials of $\nu_\mu$ and $\nu_\tau$ 
\cite{BLM}) due to the specific way the neutrino evolution equation depends 
on matter density \cite{FK}. Indeed, it was shown in Refs. \cite{KP,MW}
that $P_{ee}$ is independent of the phase $\delta_{CP}$; from \eq~(\ref{S8}) 
it then immediately follows that $P(\nu_e\to \nu_e)_{\rm dir}=
P(\nu_e\to \nu_e)_{\rm rev}$. It is interesting to note that this result
depends crucially on the assumption that the matter-induced potentials of 
all neutrino species except $\nu_e$ are the same. This is no longer true
if there are sterile neutrinos, and so in that case in general $P(\nu_e\to 
\nu_e)_{\rm dir}\ne P(\nu_e\to \nu_e)_{\rm rev}$. 

An important point to notice is that there is no T violation (either 
fundamental or matter-induced) in two-flavour neutrino systems. 
Mathematically, this
follows from the fact that the off-diagonal elements of any $2\times 2$ 
unitary matrix have the same absolute values -- for this reason the 
probabilities of the transitions $\nu_a\to\nu_b$ and $\nu_b\to\nu_a$, 
which are given by the squares of the moduli of the elements $S_{21}$ and
$S_{12}$ of the evolution matrix $S$, coincide. Physically, this is related 
to the fact that conservation of probability (\ie, unitarity) puts rather 
stringent constraints in the two-flavour neutrino case. For example, in
the $(\nu_e,\nu_\mu)$ system the conditions that the probabilities of the
transitions from $\nu_e$ to all states (including $\nu_e$ itself) and from 
all states to $\nu_e$ must both be equal to unity are
\be
P_{ee}+P_{e\mu}=1\,,\qquad
P_{ee}+P_{\mu e}=1\,.
\label{unit2}
\ee
{}From this one immediately obtains that
\be
P_{e\mu}=P_{\mu e}\,,
\label{Tsym2}
\ee
\ie, neutrino oscillations are T-invariant irrespective of whether they take 
place in vacuum or in matter and whether the matter density profile 
is symmetric or not. 

Another consequence of unitarity in the two-flavour neutrino systems is
that in the case of a symmetric matter density profile the
off-diagonal elements of the evolution matrix $S$ 
are pure imaginary. This is well known in the case of constant matter density, 
and it is easy to see that this in fact also holds for an arbitrary symmetric 
matter density profile. Indeed, in the two-flavour case the most general form 
of the unitary evolution matrix $S$ (up to phase factors which can be 
absorbed into redefinitions of the fields) is 
\be
S=\left(\begin{array}{cc}
\alpha   &  \beta \\
-\beta^* &  \alpha^*
\end{array}\right)\,, \qquad \mbox{where} \quad |\alpha|^2+|\beta|^2=1\,. 
\label{SS}
\ee
{}From \eq~(\ref{S7}) it follows that when $\tilde{N}(t)=N(t)$, the
evolution matrix is symmetric, \ie, $\beta$ is pure imaginary. 

Unitarity is much less constraining in the case of more than two neutrino 
flavours. For example, for three neutrino flavours one obtains from
equalities similar to
\eq~(\ref{unit2}) only the condition 
\be
\Delta P_{e\mu}^T + \Delta P_{e\tau}^T = 0\,,
\label{unit3}
\ee
and the T-odd asymmetries $\Delta P_{e\mu}^T$ and $\Delta P_{e\tau}^T$
need not vanish. Considering different initial neutrino states, one can
also find $\Delta P_{\mu\tau}^T+\Delta P_{\mu e}^T=0$ and 
$\Delta P_{\tau e}^T + \Delta P_{\tau\mu}^T = 0$, which together with 
\eq~(\ref{unit3}) give \cite{KrPe}
\be
\Delta P_{e\mu}^T=\Delta P_{\mu\tau}^T = \Delta P_{\tau e}^T\,.
\label{Tasym}
\ee 
This relation coincides with the corresponding relation in the case of
neutrino oscillations in vacuum; it implies that in a three-flavour
neutrino system there is only one independent T-odd probability difference 
for neutrinos (and similarly one for antineutrinos). 
In a few recent publications \cite{HS,MTKY}, this relation was obtained 
(in some approximations) for neutrino oscillations in matter of constant
density. However, as should be clear from its derivation, it is exact and 
does not depend on the matter density profile. 

It is well known that in the limits of vanishing mixing angle $\theta_{13}$ 
(\ie, vanishing element $U_{e3}$ of the lepton mixing matrix $U$) or vanishing 
mass squared difference $\Delta m_{21}^2$, three-flavour neutrino oscillations 
effectively reduce to the two-flavour ones \cite{lec}. Since, as was 
discussed above, there are no T-violating effects in the two-flavour
neutrino case, any such effect can be considered as a measure of the genuine
three-flavourness of the neutrino system. 
Studying T-violating effects in neutrino oscillations in matter can thus, 
in principle, provide us with important complementary means of measuring the 
parameters $\theta_{13}$ and $\Delta m_{21}^2$, even in the absence of 
fundamental CP and T violation.

\section{Approximate analytic description for an arbitrary matter density
profile}
\label{sec:approx}

We will give here a simple approximate expression for T-violating
effects in neutrino oscillations in a matter with an arbitrary matter density
profile. This will enable us to study the interplay of the fundamental 
and matter-induced T-violating effects in neutrino oscillations. 

We will consider a three-flavour neutrino system ($\nu_e, \nu_\mu, \nu_\tau$) 
and use the parameterization of the leptonic mixing matrix $U$ which
coincides with the standard parameterization of the quark mixing
matrix \cite{PDG}.
The leptonic mixing angle $\theta_{13}$ is constrained by the
CHOOZ reactor neutrino experiment and is known to be small
\cite{chooz}:
$$
\sin^2 2\theta_{13} \lesssim 0.10 \quad \Rightarrow \quad \sin
\theta_{13} \lesssim 0.16 \,.
$$
Analyses of solar and atmospheric neutrino data \cite{recent} also show 
that the ratio $\Delta m_{\odot}^2/\Delta m_{\rm atm}^2 \equiv \Delta
m_{21}^2/\Delta m_{31}^2 \lesssim 0.1$. As was discussed above,
T-violating effects (both fundamental and matter-induced) disappear in
the limit of vanishing $\theta_{13}$ or $\Delta m_{21}^2$ and therefore they 
must be suppressed by both these small factors. This means that they can
be calculated to a very good accuracy within perturbation theory. One can 
either consider $\Delta m_{21}^2/\Delta m_{31}^2$ as a small parameter and 
treat $\theta_{13}$ exactly, or vice versa; the T-odd asymmetry will 
automatically be suppressed by both these parameters. We choose to treat 
$\sin \theta_{13}$ as a small parameter, while making no further
approximations. The details of our calculations are described in
Appendix~A; here we just describe the main idea and the results.

In the standard parameterization, the leptonic mixing matrix $U$ can
be written
in the form $U=O_{23} V_{13} O_{12}$, where $O_{ij}$ are orthogonal 
matrices describing rotations by the angles $\theta_{ij}$ in the 
corresponding $(i,j)$-planes, and $V_{13}$ is the unitary matrix which
describes the rotation in the (1,3)-plane and in addition includes the
Dirac-type CP and T-violating phase $\delta_{CP}$ [see \eq~(\ref{stand})]. 
In the following, we will use the notation $s_{ij}\equiv\sin\theta_{ij}$, 
$c_{ij} \equiv \cos\theta_{ij}$, 
\be
\delta \equiv \frac{\Delta m_{21}^2}{2E}\,, \qquad 
\Delta \equiv \frac{\Delta m_{31}^2}{2E}\,, \qquad
\tilde{\Delta}(t) \equiv \Delta-\frac{1}{2}[\delta+V(t)]\,.
\label{deltas}
\ee
It is convenient to perform the rotation according to 
$\nu'=O_{23}^T\nu$, where $\nu$ is the neutrino vector of state in
flavour basis. To zeroth order in perturbation theory in the small
parameter $s_{13}\equiv \sin\theta_{13}$, the evolution 
matrix in the rotated basis can be written as 
\be
S_0'(t,t_0)=\left(\begin{array}{ccc}
\alpha(t,t_0)   &  \beta(t,t_0)    & ~~0 \\
-\beta^*(t,t_0)  &  \alpha^*(t,t_0) & ~~0 \\
0 & 0 & f(t,t_0)
\end{array}\right)\,,\quad\quad |\alpha|^2+|\beta|^2=1\,, 
\label{S0}
\ee
where  
\be
f(t,t_0) = \exp \left\{-i\int_{t_0}^t \tilde{\Delta}(t')\,dt'\right\}
\label{f}
\ee
and the parameters $\alpha(t,t_0)$ and $\beta(t,t_0)$ are to be determined 
from the solutions of the two-flavour neutrino problem 
in the (1,2)-subsector. It is now easy to obtain the 
correction to the evolution matrix $S_0'$ to order $s_{13}$, from which, upon 
rotation back to the unprimed basis, one finds  the following
expression for the T-odd probability difference $\Delta P_{e\mu}^T
=\Delta P_{\mu\tau}^T = \Delta P_{\tau e}^T$~\footnote{\eqs~(\ref{DeltaP}) and
(\ref{AaCa}) were obtained neglecting the (very small) corrections of the
order of $(\Delta m_{21}^2/\Delta m_{31}^2)^2$. Expressions which do not
use this approximation are given in Appendix~A.}: 
\be
\Delta P_{e\mu}^T \simeq
-2s_{13}s_{23}c_{23}\,(\Delta-s_{12}^2\delta)\, {\rm Im \,}
[e^{-i\delta_{CP}}\beta^*(A_a-C_a^*)]\,,
\label{DeltaP}
\ee
where 
\begin{equation}
A_a \equiv
\alpha\int_{t_0}^t\alpha^*f\,dt'+
\beta\int_{t_0}^t\beta^*f\,dt'\,,\qquad
C_a \equiv f\int_{t_0}^t\alpha f^*\,dt'\,.
\label{AaCa}
\end{equation}
\eqs~(\ref{DeltaP}) and (\ref{AaCa}) give the T-odd probability 
difference in the three-flavour neutrino system in terms of $f$ and the
two-flavour neutrino amplitudes $\alpha$ and $\beta$ for an arbitrary
matter density profile. Note that this simplifies the problem
considerably as the two-flavour neutrino problems are generally much
easier to solve than the three-flavour ones. For $\delta_{CP} = 0$ it
can be shown that the right-hand side of \eq~(\ref{DeltaP}) vanishes
for any symmetric matter density profile, as it must (see Appendix~A). 

We will now consider two special cases: first, matter consisting of
two layers of constant densities, and second, the adiabatic
approximation for an arbitrary matter density profile. 
For the first case with two layers of constant densities we will use the 
following notation: layer widths $L_1$ and $L_2$, electron number
densities $N_1$ and $N_2$, the corresponding matter-induced potentials $V_1$ 
and $V_2$, the values of the mixing angle in the (1,2)-subsector 
in matter $\theta_1$ and $\theta_2$
\footnote{In the limit of vanishing densities $N_1$ and $N_2$, one has 
$\theta_1=\theta_2=\theta_{12}$.}, respectively, 
\be
\omega_{i} \equiv \frac{1}{2}\sqrt{(\cos 2\theta_{12}\,\delta-V_{i})^2+
\sin^2 2\theta_{12}\,\delta^2}\,,
\label{omega}
\ee
and 
\be
s_i \equiv \sin(\omega_i L_i)\,, \qquad c_i \equiv\cos(\omega_i
L_i)\,, \qquad (i = 1,2).
\label{sc}
\ee
For the T-odd difference of the neutrino oscillation probabilities one
then finds 
\be
\Delta P_{e\mu}^T\simeq -2s_{13}s_{23}c_{23}\,(\Delta-s_{12}^2\delta)\,
[\cos\delta_{CP}(X_1 Z_R-X_2 Z_I)+\sin\delta_{CP}(X_1 Z_I+X_2 Z_R)]\,.
\label{DeltaPP}
\ee
Here $X_1$, $X_2$, $X_3$, and $Y$ are the parameters which define the 
two-flavour evolution matrix for neutrinos passing through the two-layer 
medium, $s=Y-i\mbox{\boldmath{$\sigma$}} \cdot \bf{X}$, with \cite{Akh2}
\begin{eqnarray}
& & Y = c_1 c_2-s_1 s_2\cos(2\theta_1-2\theta_2)\,,\qquad
X_1 = s_1 c_2\sin 2\theta_1+s_2 c_1\sin 2\theta_2\,, \nonumber\\
& & X_2 = -s_1 s_2 \sin(2\theta_1-2\theta_2)\,,\qquad
X_3 = -(s_1 c_2\cos 2\theta_1+s_2 c_1\cos 2\theta_2)\,, \nonumber
\end{eqnarray}
and 
\begin{eqnarray}
Z_R &\equiv& {\rm Re \,} (A_a-C_a^*) = -[D_{-}-(\Omega_1\cos 2\theta_1 -
\Omega_2\cos 2\theta_2)] \nonumber \\
&\times&  [\sin(\Delta_1 L_1 +\Delta_2 L_2) 
-c_2\sin(\Delta_1 L_1)-c_1\sin(\Delta_2 L_2)]
\nonumber \\
&-& [D_{-}\cos 2\theta_1+\Omega_2\cos(2\theta_1-2\theta_2)-\Omega_1]
s_1\cos(\Delta_2 L_2) 
\nonumber \\
&-& [D_{-}\cos 2\theta_2-\Omega_1\cos(2\theta_1-2\theta_2)+\Omega_2]
s_2\cos(\Delta_1 L_1)
\nonumber \\
&-& D_{-}X_3-\Omega_1[s_1 c_2+s_2 c_1\cos(2\theta_1-2\theta_2)] 
+ \Omega_2[c_1 s_2+s_1 c_2\cos(2\theta_1-2\theta_2)]\,,
\label{ZR}
\end{eqnarray}
\begin{eqnarray}
Z_I &\equiv& {\rm Im \,} (A_a-C_a^*) = [D_{+}-(\Omega_1\cos 2\theta_1 + 
\Omega_2\cos 2\theta_2)] \cos(\Delta_1 L_1 +\Delta_2 L_2) 
\nonumber \\
&-& [D_{-}-(\Omega_1\cos 2\theta_1-\Omega_2\cos 2\theta_2)]
[c_1\cos(\Delta_2 L_2)-c_2\cos(\Delta_1 L_1)]
\nonumber \\
&-& [D_{-}\cos 2\theta_1+\Omega_2\cos(2\theta_1-2\theta_2)-\Omega_1]
s_1\sin(\Delta_2 L_2) 
\nonumber \\
&+& [D_{-}\cos 2\theta_2-\Omega_1\cos(2\theta_1-2\theta_2)+\Omega_2]
s_2\sin(\Delta_1 L_1) 
\nonumber \\
&-& D_{+}Y+\Omega_1(c_1 c_2\cos 2\theta_1-s_1 s_2\cos 2\theta_2)
+\Omega_2(c_1 c_2\cos 2\theta_2-s_1 s_2\cos 2\theta_1)\,.
\label{ZI}
\end{eqnarray}
Here we have used the notation 
\begin{eqnarray}
\Delta_i = \Delta-\frac{1}{2}(\delta+V_i)\,, \quad
D_{\pm} = \frac{\Delta_1}{\Delta_1^2-\omega_1^2}
\pm \frac{\Delta_2}{\Delta_2^2-\omega_2^2}\,, \quad
\Omega_i = \frac{\omega_i}{\Delta_i^2-\omega_i^2}\, \quad (i = 1,2).
\end{eqnarray}
The term proportional to $\cos\delta_{CP}$ in 
\eq~(\ref{DeltaPP}) describes the matter-induced T violation, whereas
the $\sin\delta_{CP}$ term is due to the fundamental T violation.  
It is easy to see that the parameters $X_2$ and $Z_R$ are antisymmetric with 
respect to the interchange of the two layers, while $X_1$, 
$X_3$, $Y$, and $Z_I$ are symmetric, and therefore the condition (\ref{S8}) is 
satisfied. 

In the low energy regime $\delta=\Delta m_{21}^2/2E \gtrsim V_{1,2}$
(which for the LMA-MSW solution of the solar neutrino problem and matter
densities typical for the upper mantle of the earth corresponds to $E
\lesssim 1$ GeV) the main contributions to \eq~(\ref{DeltaPP}) come 
from the $D_{+}$ terms in $Z_I$, and the expression for the T-odd 
probability difference simplifies significantly: 
\begin{eqnarray}
\Delta P_{e\mu}^T &\simeq& \cos\delta_{CP}\cdot 8s_{12} c_{12} s_{13} s_{23} 
c_{23}\,\frac{\sin(2\theta_1-2\theta_2)}{\sin 2\theta_{12}}\{s_1 s_2[Y-
\cos(\Delta_1 L_1+\Delta_2 L_2)]\} \nonumber \\
&+& \sin\delta_{CP}\cdot 4 s_{13}s_{23}c_{23} \, 
X_1[Y-\cos(\Delta_1 L_1+\Delta_2 L_2)] \,.
\label{DP1}
\end{eqnarray}
Here the $\cos\delta_{CP}$ term has a remarkably simple structure: it is
given by an oscillating term multiplied by an effective Jarlskog invariant
\be
J_{\rm eff} \equiv s_{12} c_{12} s_{13}s_{23}c_{23}\,\frac{\sin(2\theta_1-
2\theta_2)}{\sin 2\theta_{12}}\,.
\label{Jeff}
\ee
This has to be compared with the usual Jarlskog invariant
\cite{Jarl}
\be
J \equiv s_{12} c_{12} s_{13} c_{13}^2 s_{23}c_{23}\,\sin\delta_{CP}\,.
\label{J}
\ee
Note that the factor $\sin(2\theta_1-2\theta_2)/\sin 2\theta_{12}$ in
$J_{\rm eff}$ plays the same role as $\sin\delta_{CP}$ in $J$ (which
is zero in the absence of fundamental CP and T violation). The factor
$\sin(2\theta_1-2\theta_2)/\sin 2\theta_{12}$ is a measure of 
the asymmetry of the matter density profile ($\theta_1 \ne \theta_2$). 
Note also the absence of the $c_{13}^2$ factor which is
equal to unity in our approximation. The $\sin\delta_{CP}$ term in
\eq~(\ref{DP1}) can also be expressed in terms of the (usual) Jarlskog
invariant if one writes $X_1$ as
\be
X_1=\sin 2\theta_{12}
\left(s_1 c_2\frac{\sin 2\theta_1}{\sin 2\theta_{12}}+s_2 c_1
\frac{\sin 2\theta_2}{\sin 2\theta_{12}}\right)\,.
\label{X1new}
\ee
Note that the ratios $\sin 2\theta_1/\sin 2\theta_{12}$ and $\sin 2\theta_2
/\sin 2\theta_{12}$ are finite in the limit $\sin 2\theta_{12}\to 0$.

Let us now consider an arbitrary matter density profile in the adiabatic 
approximation (see Appendix~B). In this case the parameters
$\alpha$ and $\beta$ describing the two-flavour neutrino evolution in
the (1,2)-subsector are
\begin{eqnarray}
\alpha(t,t_0) &=& \cos\Phi\cos(\theta-\theta_0)+i\sin\Phi\cos(\theta+
\theta_0)\,, \label{alphabeta2}\\
\beta(t,t_0) &=& \cos\Phi\sin(\theta-\theta_0)-i\sin\Phi\sin(\theta+
\theta_0)\,, 
\label{alphabeta}
\end{eqnarray}
whereas $f(t,t_0)$ is given by \eq~(\ref{f}) as before.
Here $\theta_0$ and $\theta$ are the values of the mixing angle in the
(1,2)-subsector in matter at the initial and final points of the neutrino
evolution, $t_0$ and $t$, respectively, 
\be
\Phi \equiv \int_{t_0}^t\omega(t')\,dt'\,,
\label{Phi}
\ee
and $\omega(t)$ is given by \eq~(\ref{omega}) with the substitution 
$V_{1,2}\to V(t)$. 
We will also assume that $V(t) \lesssim \delta\ll \Delta$. The integrals 
in \eq~(\ref{AaCa}) can then be done approximately (see
Appendix~B) and one obtains
\begin{eqnarray}
\Delta P_{e\mu}^T\simeq 2s_{13}s_{23}c_{23}\,\{\cos\delta_{CP}
[\sin(2\theta-2\theta_0)\cos^2 \Phi-2\sin(\theta-\theta_0)\cos\Phi
\cos(\Delta(t-t_0))]+ \nonumber \\
\sin\delta_{CP}[\sin(\theta+\theta_0)\cos(\theta-\theta_0)\sin 2\Phi-
2\sin(\theta+\theta_0)\sin\Phi\cos(\Delta(t-t_0))]\} \,.
\label{DP2}
\end{eqnarray}
In the regime in which the oscillations governed by large $\Delta =
\Delta m_{31}^2/2E$ are fast and therefore can be averaged over, the above
expression simplifies to 
\be
\Delta P_{e\mu}^T\simeq 2s_{13}s_{23}c_{23}\,\{\cos\delta_{CP} 
\sin(2\theta-2\theta_0)\cos^2 \Phi+
\sin\delta_{CP}\sin(\theta+\theta_0)\cos(\theta-\theta_0)\sin 2\Phi\} \,.
\label{DP3}
\ee
Note that the $\cos\delta_{CP}$ contribution can again be written in
terms of the effective Jarlskog invariant $J_{\rm eff}$: 
\be
(\Delta P_{e\mu}^T)_{\cos\delta_{CP}} \simeq 4\cos\delta_{CP}\cdot
s_{12}c_{12}s_{13} s_{23}c_{23}\frac{\sin(2\theta-2\theta_0)}{\sin
2\theta_{12}}\cos^2\Phi=4 \cos\delta_{CP}\,J_{\rm eff}\cos^2\Phi\,.
\label{DP4}
\ee
Interestingly, the effective Jarlskog invariant appears both in 
the adiabatic regime and in the case of the two-layer matter density
profile, which is an example of an extreme non-adiabatic case. 
If the oscillations due to a smaller energy difference
$\omega$ are also in the averaging regime, the $\sin\delta_{CP}$
contribution in \eq~(\ref{DP3}) vanishes and only the matter-induced
T-violating term proportional to $\cos\delta_{CP}$ survives. Thus, we
make the interesting observation that
matter-induced T violation, unlike the fundamental one, does not
disappear when the neutrino oscillations are in the regime of complete
averaging.

\section{Implications }
\label{sec:impl}

In this section we compare our approximate analytic formulas with 
numerical calculations of $\Delta P_{e\mu}^T$ and discuss the relevance
of matter-induced T violation in neutrino oscillations for long baseline 
experiments, and also for solar, atmospheric and supernova neutrinos and
for neutrinos in the early universe.

\subsection{Accuracy of the analytic approximation}

\begin{figure}[htpb!]
  \begin{center}
    \includegraphics[width=0.3 \textwidth,angle=-90]{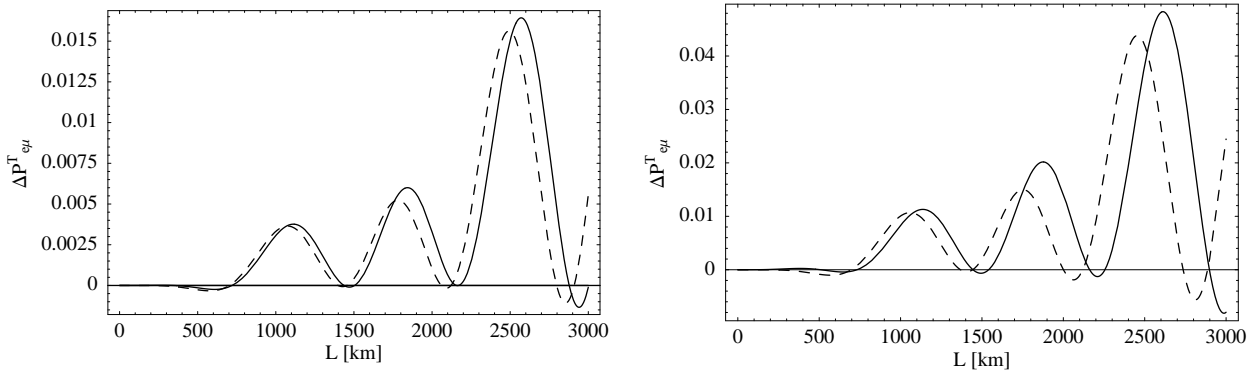}
\caption{\small Comparison of the results of the analytic (solid curve) 
and numerical (dashed curve) calculations of $\Delta P_{e\mu}^T$ for two 
layers of widths $L_1=L_2=L/2$, densities 1 g/cm$^3$ and 3 g/cm$^3$ 
and electron number fractions $(Y_e)_1=(Y_e)_2=0.5$.  
In both plots $E=1$ GeV, $\Delta m^2_{31}=3.5\cdot10^{-3}$ 
eV$^2$, $\delta_{CP}=0$, $\theta_{12}=0.56$, and $\theta_{23}=\pi/4$
are used. The remaining parameters are $\theta_{13}=0.1$ and 
$\Delta m^2_{21}=5\cdot10^{-5}$ eV$^2$ for the left plot and 
$\theta_{13}=0.16$ and $\Delta m^2_{21}=2\cdot 10^{-4}$ eV$^2$ for the 
right plot.} 
    \label{fig:compPhase}
  \end{center}
\end{figure}

In order to estimate the accuracy and domain of validity of our results we 
have to distinguish two cases: $L/E \lesssim 10,000$ km/GeV and $L/E
\gtrsim 10,000$ km/GeV. In 
the first case the oscillating structure of $\Delta P_{ab}^T$ can, in
general, be resolved, as it is shown in \fig~\ref{fig:compPhase}. As
can be seen from the figure, the size of $\Delta P_{e\mu}^T$ is
reproduced rather accurately, but there is an error in the oscillation
phase which accumulates with distance and increases with increasing
$\theta_{13}$ and $\Delta m^2_{21}$.

\begin{figure}[htpb!]
  \begin{center}
    \includegraphics[width=0.3 \textwidth,angle=-90]{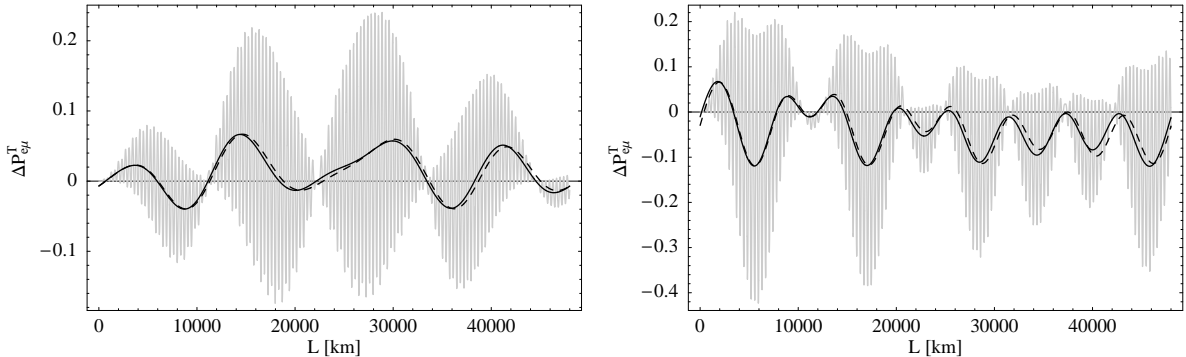}
 \caption{\small Comparison of the analytic and
numerical calculations of $\Delta P_{e\mu}^T$
for two layers of widths $L_1=L_2=L/2$, densities $0$ and 6.4 g/cm$^3$ 
and electron number fractions $(Y_e)_1=(Y_e)_2=0.5$.  
The grey curve is the result of our analytic calculation, the 
black solid and dashed curves show the results averaged over the fast
oscillations of the analytic and numerical calculation, respectively. 
In both plots $E=0.5$ GeV, $\Delta m^2_{31}=3.5\cdot10^{-3}$ 
eV$^2$, $\delta_{CP}=0$, $\theta_{12}=0.56$, and $\theta_{23}=\pi/4$
are used. The remaining parameters are $\theta_{13}=0.1$ and $\Delta
m^2_{21}=5\cdot10^{-5}$ 
eV$^2$ for the left plot and $\theta_{13}=0.16$ and $\Delta m^2_{21}=
2\cdot10^{-4}$ eV$^2$ for the right plot.} 
    \label{fig:compAverage}
  \end{center}
\end{figure}

In the second case, the oscillations governed by the large $\Delta m_{31}^2=
\Delta m_{\rm atm}^2$ are very fast. We illustrate this case in
\fig~\ref{fig:compAverage}, 
where $\Delta P_{e\mu}^T$ is plotted as a function of the total distance
$L$ traveled by neutrinos for matter consisting of two layers with densities 
$\rho_1=0$, $\rho_2=6.4$ g/cm$^{3}$ and widths $L_1=L_2=L/2$. We have
chosen for the left plot the same parameters as those in
\fig~3a of \Ref \cite{FK}
\footnote{In \Ref \cite{FK} the value $\rho_2=8$ g/cm$^3$ instead of 
$6.4$ g/cm$^3$ was erroneously quoted for \fig~3a, but the quoted
value of the potential $V_2$ was correct. Note that in \fig~3a of 
\Ref \cite{FK} the asymmetry of the transition probability $P_{\mu e}$ with 
respect to density profile reversal rather than $\Delta P_{e\mu}^T$ was
plotted. However, as follows from \eq~(\ref{S7}), in the case of 
$\delta_{CP}=0$ this asymmetry coincides with $\Delta P_{\mu e}^T=-\Delta
P_{e\mu}^T$.}. The baseline values are rather unrealistic and serve
illustrative purposes only. In the right plot of
\fig~\ref{fig:compAverage} we plot $\Delta P_{e\mu}^T$
for larger values of $\theta_{13}$ and $\Delta m_{21}^2$. 
The result obtained using our analytic formula (\ref{DeltaPP})
(grey curve in \fig~\ref{fig:compAverage}) reproduces the results of 
numerical calculations of \Ref \cite{FK} and of our own numerical calculations
very well: the difference can be barely seen. In fact, the fast oscillations 
cannot be resolved by any realistic detector due to its limited energy 
resolution; for this reason in \fig~\ref{fig:compAverage} we compare the 
averaged over these fast oscillations values of $\Delta P_{e\mu}^T$ obtained 
using approximate analytic formula (\ref{DeltaPP}) with those calculated 
numerically (black solid and dashed curves, respectively). Since the error
in the phase is irrelevant due to the averaging, the accuracy of the analytic 
approach is very good. For $\theta_{13}=0.1$ and $\Delta m^2_{21}=
5\cdot10^{-5}$ eV$^2$ the maximal relative error is about 5\%; it increases 
with increasing $\theta_{13}$ and $\Delta m^2_{21}$, and for 
$\theta_{13}=0.16$ and $\Delta m^2_{21}=2\cdot10^{-4}$ eV$^2$ the maximal 
error is about 10\%.

In both cases the accuracy of the predicted size of matter-induced T
violation is quite good, the error being $\lesssim 10\%$; however, the phase 
is not reproduced to the same accuracy. 

\subsection{Long baseline experiments}

One possible implication of the discussed effects may be in future long 
baseline neutrino oscillation experiments. Neutrino factory
experiments with thousands of kilometers baselines appear feasible, leading
to beams which traverse the matter of the earth. 
The matter density profile of the earth is approximately spherically
symmetric;  
however, the deviations from exact spherical symmetry may be considerable, 
which in turn can give rise to T-violating matter density profiles seen by
neutrino beams. These fine details of the earth structure are, unfortunately, 
not very well known, but from seismological data one can infer some upper
bounds on the possible asymmetry. 
A conservative estimate of these deviations is to assume a 10\% 
variation in density on the length scales of several thousand
kilometers \cite{Igel} somewhere along the neutrino path.
Another possibility to have an asymmetric matter density profile exists for 
shorter baselines $L\lesssim 1,000$ km where the neutrino path goes only some
ten kilometers below the surface of the earth. In this case one half of the 
neutrino path could be in the sea ($\rho \simeq 1$ g/cm$^3$) and the other 
half in the continental earth crust ($\rho \simeq 3$ g/cm$^3$). The
experimental setups under discussion will have a remarkable precision,
and it is thus 
important to know when the discussed matter-induced T-violating
effects become relevant. In other words, we will discuss when it is
necessary to include the discussed extra T-violating effects and when
the usual analysis is justified.

\begin{figure}[hb!]
  \begin{center}
    \includegraphics[width=0.33 \textwidth,angle=-90]{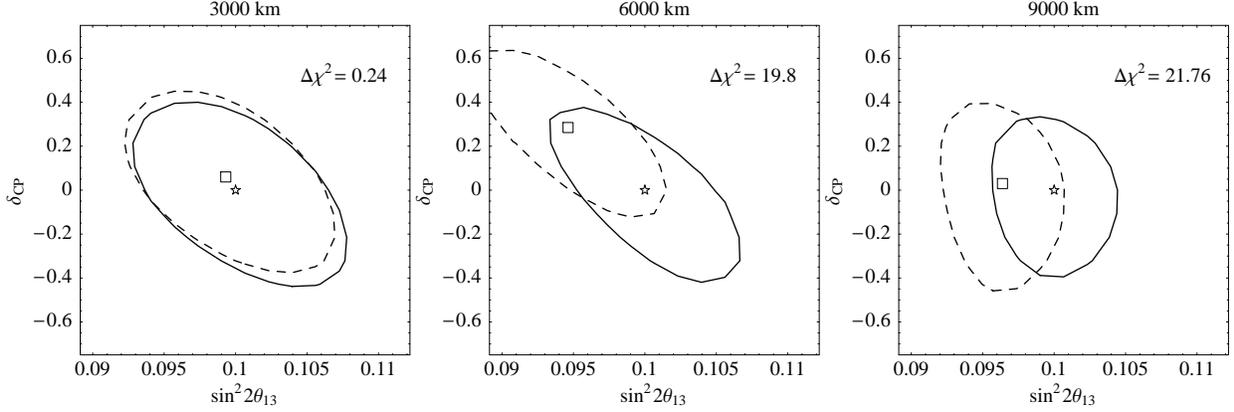}
    \caption{\small A fit of $\sin^2 2\theta_{13}$ and
      $\delta_{CP}$ for different baselines in a neutrino factory
      experiment with a beam energy of 50 GeV (see the text for details).
The solid curve is the 99\% C.L. 
contour for the asymmetric matter density profile with two  
layers of widths $L_1=L_2=L/2$, densities 1 g/cm$^3$ and 3~g/cm$^3$ 
and electron number fractions $(Y_e)_1=(Y_e)_2=0.5$.  
The dashed curve is the 99\% C.L. contour for the 
symmetrized matter density profile as defined in \eq~(\ref{eq:symProfile}). 
The star and square indicate the best fit points for the asymmetric and 
symmetrized profiles, respectively. The $\Delta\chi^2$ value given in each 
plot is the difference of $\chi^2$ (2 d.o.f.) between the two best fit
points. The parameters are $\theta_{12}=\pi/4$, $\theta_{23}=\pi/4$, 
$\Delta m^2_{31}=3.5\cdot10^{-3}$ eV$^2$, $\Delta m^2_{21}=10^{-4}$ eV$^2$, 
$\sin^2 2\theta_{13}=0.1$, and $\delta_{CP}=0$.}
    \label{fig:stepEggs}
  \end{center}
\end{figure}

\begin{figure}[htpb!]
  \begin{center}
    \includegraphics[width=0.33 \textwidth,angle=-90]{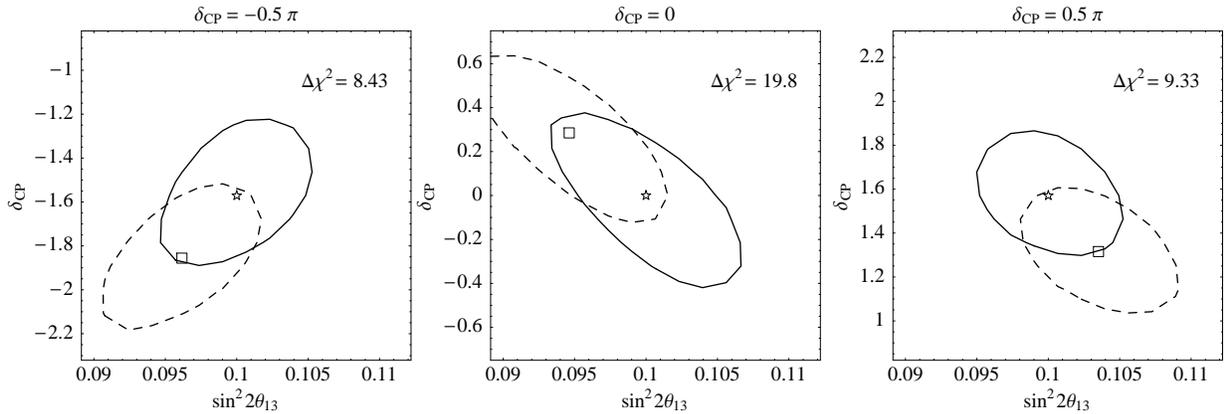}
\caption{\small Same as in \fig~\ref{fig:stepEggs}, but for the fixed
baseline $L=6,000$ km and three different values of $\delta_{CP}$:
$-\pi/2$, 0, and $\pi/2$.}
   \label{fig:stepEggsCP}
  \end{center}
\end{figure}

We consider the usual setup for neutrino factory experiments with a beam 
energy of 50 GeV. The energy threshold of the detector is 4 GeV and its
energy resolution is 10\%. The detector is capable of charge identification, 
thus the wrong sign muon signal is available. We assume $2\cdot10^{21}$
useful muon decays for both polarities and a detector mass of 40 kton. 
This luminosity coincides with the one used in \Ref \cite{Cervera} and is
a factor of 40 larger than the one used in \Ref \cite{Freund:2000ti}. We
include only statistical errors. All fits include both polarities and both 
appearance and disappearance rates. Further details can be found in 
\Ref \cite{Freund:2000ti}.

\begin{figure}[htpb!]
  \begin{center}
    \includegraphics[width=0.33 \textwidth,angle=-90]{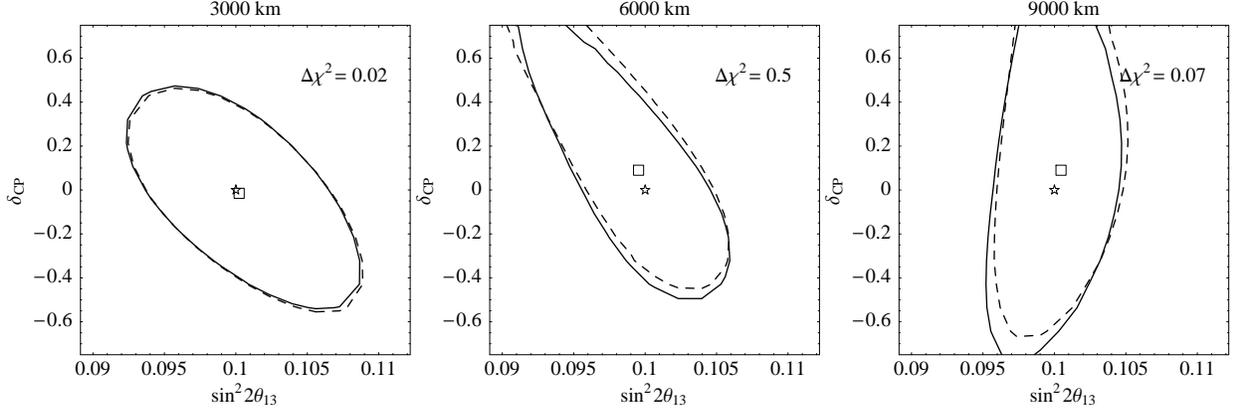}
\caption{\small Same as in \fig~\ref{fig:stepEggs}, but for the layer 
densities $\rho_1=3$ g/cm$^3$ and $\rho_2=3.3$ g/cm$^3$. 
}
    \label{fig:realisticEggs}
  \end{center}
\end{figure}

We assume two different asymmetric matter density profiles. The first
one consists of two layers of equal widths, with densities 1 g/cm$^3$
and 3 g/cm$^{3}$ (\figs~\ref{fig:stepEggs} and \ref{fig:stepEggsCP}).
This
corresponds to the sea-earth scenario, which of course can be realized on
the earth only up to baselines of $\sim 1,000$ km. The second one
also consists of two layers of equal widths, but with densities 3 g/cm$^3$
and 3.3 g/cm$^{3}$ as an example of density perturbations which could arise 
in real very long baseline experiments (\fig~\ref{fig:realisticEggs}).

We simulate such experiments numerically for these types of profiles
and perform fits to the obtained event rates. We compare this with the
fits performed for symmetrized versions of the corresponding
profiles, which are modeled by replacing the transition probabilities
by the symmetrized ones:
\begin{equation}\label{eq:symProfile}
P_{S}=\frac{1}{2}\left(P_{\rm dir}+P_{\rm rev}\right)\,.
\end{equation}
Thus we are only sensitive to the errors induced by the asymmetry of 
the matter density profile and not to the errors in the average density, which 
of course should not be ignored in reality. The difference between the
minimal $\chi^2$ values for the asymmetric and symmetrized profiles is
a direct measure of the sensitivity to the matter-induced T violation. 

The effect can be quite sizable for matter density profiles of the
sea-earth type (see \fig~\ref{fig:stepEggs}), however, only for 
baselines above 1,000 km, which cannot be realized on the earth. For
large baselines the errors in the determination of the fundamental CP
and T-violating phase $\delta_{CP}$ induced by asymmetric matter are
comparable with the statistical errors in the case of symmetric matter. 
This behavior is quite similar for all
possible values of $\delta_{CP}$ as can be seen in
\fig~\ref{fig:stepEggsCP}. 

\begin{figure}[htpb!]
  \begin{center}
    \includegraphics[width=0.3 \textwidth,angle=-90]{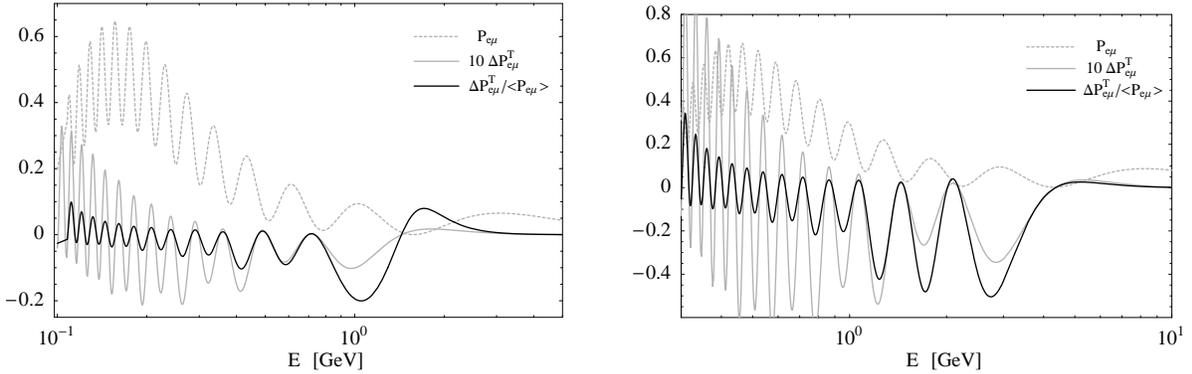}
     \caption{\small Energy dependence of the transition probability 
$P_{e\mu}$ (grey dashed curve), T-odd probability difference $\Delta
P_{e\mu}^T$ multiplied by 10 (grey solid curve), and the ratio 
$\Delta P_{e\mu}^T/\langle P_{e\mu}\rangle$, where $\langle P_{e\mu}\rangle$ 
is the average of $P_{e\mu}$ over fast oscillations (black solid curve) 
in the case of two-layer density profile with $\rho_1=1$ g/cm$^3$, 
$\rho_2=3$ g/cm$^3$ and electron number fractions $(Y_{e})_1=(Y_{e})_2=0.5$. 
The layer widths are $L_1=L_2=500$ km (left plot) and $L_1=L_2=1,500$ km
(right plot). The values of neutrino parameters are
$\Delta m_{21}^2=2\cdot 10^{-4}$ eV$^2$, $\theta_{12}=0.559$, 
$\Delta m_{31}^2=3.5\cdot 10^{-3}$ eV$^2$, $\theta_{23}=\pi/4$, 
$\theta_{13}=0.16$, and $\delta_{CP}=0$.} 
    \label{fig:P_ratio}
  \end{center}
\end{figure}

The energy dependence of the transition probability $P_{e\mu}$, T-odd
probability difference $\Delta P_{e\mu}^T$, and their ratio in the case of 
two-layer matter density profile corresponding to the sea-earth
scenario is shown  
in \fig~\ref{fig:P_ratio}. One can see that the relative size of the 
T-violating effect is largest at energies of about 1 GeV. A dedicated 
low energy experiment capable of measuring the matter-induced T violation
with the sea-earth type matter density profile at a baseline of $\sim 1,000$ km
would, however, require enormous luminosities, at least six orders of
magnitudes higher than an initial stage of a neutrino factory ($2
\cdot 10^{19}$ muons/year, 10 kton).

For much more realistic matter density profiles with only 10\% density
variation, the matter-induced T violation effects are small at any
baseline (see \fig~\ref{fig:realisticEggs}). The statistical errors in
the determination of $\delta_{CP}$ and $\theta_{13}$ are much larger than 
the errors induced by replacing the asymmetric matter density profile by the
symmetrized one. This again holds for all values of $\delta_{CP}$.
Therefore we conclude that the determination of the fundamental CP
and T-violating phase $\delta_{CP}$ cannot be spoiled by an unknown T
asymmetry of the matter density profile until the luminosity (and therewith 
the accuracy) is increased by at least two orders of magnitude, bringing
the error in $\delta_{CP}$ down to $\sim 1$\% at 99\% C.L.

\subsection{Solar, supernova and atmospheric neutrinos and neutrinos in 
the early universe}

Matter-induced T-violating effects can, in principle, manifest themselves 
in a number of other situations when oscillating neutrinos propagate
through asymmetric matter. In particular, solar neutrinos traveling from
the center of the sun towards its surface traverse a matter density profile 
which is highly asymmetric and characterized by a large contrast of 
densities (from $\rho~\simeq 150$ g/cm$^3$ to $\rho\simeq 0$). There are 
no muon or tau neutrinos originally produced in the nuclear reactions
in the sun, and so one cannot study the quantities like $\Delta P_{e\mu}^T$ 
or $\Delta P_{e\tau}^T$ with solar neutrinos 
\footnote{Neutrinos of all flavours with energies ${\cal O}$(GeV) can be 
produced in the annihilation of weakly interacting massive particles 
(WIMPs) inside the sun.  T-violating effects of solar matter on the
oscillations of such neutrinos were discussed in \Ref~\cite{Andre}.}.
However, genuine three-flavour effects
(including those of asymmetric matter) in general contribute terms of the 
order of $s_{13}$ to the transition probabilities which can be large compared 
to the usual terms proportional to $s_{13}^2$ which are present in the 
effective two-flavour approach. In particular, the survival probability 
of solar neutrinos averaged over the fast oscillations due to the large 
$\Delta m_{\rm atm}^2$ is approximately given by ``quasi two-flavour''
formula ~\cite{Lim87} 
\be
P^{3\nu}_{ee}=c_{13}^4 P^{2\nu}_{ee}+s_{13}^4\,,
\label{surv}
\ee
where $P^{2\nu}_{ee}$ is the two-flavour $\nu_e$ survival probability in the 
(1,2) sector calculated with the effective matter-induced neutrino potential 
$V_{\rm eff}= c_{13}^2 V$. It contains only second and higher order in
$s_{13}$ corrections to the standard two-flavour result. One can then ask 
if this result is modified by asymmetric matter, leading to the order
$s_{13}$ corrections. 

In the case of the LMA-MSW solution of the solar neutrino problem (which 
leads to the largest allowed values of $\Delta m_{21}^2$) the evolution of 
neutrinos inside the sun is adiabatic, and \eqs~(\ref{DP2}) or (\ref{DP3})
can be used. For maximal values of $\theta_{13}$ allowed by the CHOOZ
experiment one finds that the corrections of the order $s_{13}$ to
the transition probabilities $P_{e\mu}$ and $P_{e\tau}$ of solar neutrinos 
can be as large as (5 -- 10)\%. However, these corrections have opposite 
sign and the same absolute value, and therefore they cancel exactly in the 
total transition (or survival) probability of electron neutrinos. The fact
that the survival probability of $\nu_e$ is symmetric with respect to matter 
density profile reversal has already been discussed above (see 
Sec. \ref{sec:general}); since one cannot distinguish $\nu_\mu$ from 
$\nu_\tau$ in low energy experiments, this makes the asymmetric matter
effects in oscillations of solar neutrinos inside the sun unobservable. 
\eq~(\ref{surv}) therefore does not have to be modified. We have checked this 
by comparing its predictions with the results of numerical calculations 
of the full three-flavour evolution equation for solar neutrinos and found
that the accuracy of the approximation in \eq~(\ref{surv}) is extremely good 
(error $\leq 10^{-3}$). 
 
Another potentially interesting implication of matter-induced 
T-violating effects could be in oscillations of supernova 
neutrinos propagating from the supernova's core outwards. However, these 
effects again cancel because of the absence of the matter-induced
asymmetry of $P_{ee}$ and the fact that the fluxes and spectra of the 
supernova $\nu_\mu$ and $\nu_\tau$ are practically identical. 

The expansion of the universe implies that neutrino oscillations in
the early universe take place in a time-dependent, asymmetric environment. 
One can then ask if T violation effects in neutrino oscillations can
result in a lepton flavour asymmetry. Such an asymmetry, \eg, in electron 
neutrinos could have important consequences for the big bang
nucleosynthesis. However, it is easy to make sure that if the original
numbers of neutrinos of all species are equal, this does not happen:  
the summation over the contributions of all flavours makes the asymmetries 
in the individual lepton flavours vanish. 

Thus, we see that even though there are, in principle, interesting extra 
T-violating effects in asymmetric situations like in the sun, in the early
universe and in supernovae explosions, these effects cancel whenever a 
summation over neutrino flavours is present, and a naive analysis without
these effects leads to correct results.

Matter-induced T violation can, in principle, also influence atmospheric
neutrino oscillations. This is related to the fact that neutrinos are 
produced in the atmosphere at an average height of about 15 km and so 
the neutrinos coming to the detector from the lower hemisphere travel
first in air and then in the earth. For neutrinos from just below the 
horizon this can result in nearly equal pathlengths (of the order of 
a few hundred kilometers) in air and in the earth. The effect is, however, very
small, which can be seen through a simple estimate. The asymmetric matter
effects can be looked for through the distortions of the zenith angle 
distributions of the $e$-like events (the distributions of the $\mu$-like 
events are mainly governed by the large $\theta_{23}$ and $\Delta
m_{31}^2$, and the relative T-violating effects for them are smaller than 
those for the $e$-like events).   
The probability difference $\Delta P_{e \mu}^T$ is of the order of a few
percent of $P_{e \mu}$, and $P_{e \mu}$ itself is of the order of a 
few percent of $P_{ee}$; therefore the matter-induced T-violating effect
is of the order $10^{-4}-10^{-3}$ at best. Thus, one would need $10^6-10^8$ 
events to achieve a statistical accuracy of the same order. Super-Kamiokande 
at the moment has roughly $10^4$ events. Including the flux uncertainties
would make the situation even worse. Therefore matter-induced T-violating
effects in atmospheric neutrinos can be safely neglected. 
This result is also confirmed by numerical analysis.

\section{Discussion and conclusions}
\label{sec:dc}
In the present paper we studied T violation in neutrino oscillations 
in asymmetric matter, with a special emphasis on matter-induced T violation.  
To this end, we derived a simple approximate analytic expression for 
T-odd probability differences $\Delta P_{ab}^T$ in the case of a general
matter density profile. We have shown that our analytic expressions 
reproduce the results of direct numerical integration of the neutrino 
evolution equation very well. 

Matter-induced T violation has two aspects. First, it is an 
interesting matter effect, which is present only in asymmetric matter 
(and therefore is absent in the most studied case of constant-density 
matter). It can manifest itself not only in the T-odd differences of 
oscillation probabilities but also in specific modification of the 
probabilities themselves.  
Since this effect can only exist in systems of three or more neutrino 
flavours, it is sensitive to the parameters that discriminate between genuine 
three-flavour and effectively two-flavour oscillations, such as $\theta_{13}$ 
and $\Delta m_{21}^2$, and therefore can, in principle, be used for their
determination. Second, matter-induced T violation can fake the
fundamental one and so impede the determination of the fundamental CP and 
T-violating phase $\delta_{CP}$ in the long baseline experiments. 

We have studied both these aspects and have shown that in the case of three 
neutrino species and for matter density contrasts and baselines feasible 
in terrestrial experiments, matter-induced 
T violation effects are small and can safely be ignored. This is due to
the fact that they are doubly suppressed by small factors $\sin\theta_{13}$ 
and $\Delta m_{21}^2/\Delta m_{31}^2=\Delta m_{\odot}^2/\Delta m_{\rm atm}^2$. 
In particular, asymmetric matter effects cannot hinder the determination of 
the fundamental CP and T-violating phase $\delta_{CP}$ in the long baseline 
experiments as far as the error in the determination of $\delta_{CP}$ is
larger than 1\% at 99\% C.L. This is the main result of our paper. 

T-asymmetric matter effects for solar, supernova, and cosmological 
neutrinos, which propagate through larger density contrasts and over larger 
distances, can, in principle, be large in each individual oscillation
channel. However, these effects are not observable because of the summation 
over channels inherent in the experimental detection of these neutrinos or
in the observable quantities such as total number density of neutrinos of a 
given flavour in the case of neutrinos in the early universe. 

The situation can be very different in four-neutrino schemes. In that case 
T-violating effects (both fundamental and matter-induced) are not, in
general, suppressed by small factors like $\sin\theta_{13}$ or 
$\Delta m_{\odot}^2/\Delta m_{\rm atm}^2$. Therefore experimental
detection of sizeable T violation effects would signify the existence of 
a fourth light neutrino, which, due to the LEP result on the invisible 
$Z^0$ boson width, must be a sterile neutrino. It should also be 
understood that our conclusion that matter-induced T violation cannot 
hamper the determination of the fundamental CP and T-violating phases 
does not, in general, apply to the four-neutrino schemes; in this case 
one has to carefully take matter-induced T-violation into account in 
order to disentangle intrinsic T violation from the extrinsic one.  

It is also interesting to note that the statement that the survival 
probability of electron neutrinos $P_{ee}$ is invariant with respect to 
matter density profile reversal, discussed in Sec. \ref{sec:general}, 
holds only in three-neutrino schemes. Indeed, it relies on the assumption 
that the matter-induced potentials of all neutrino species except $\nu_e$ 
are the same. This is no longer
true if sterile neutrinos are present. Therefore an observation of 
non-invariance of $P_{ee}$ under density profile reversal would be a
signature of a sterile neutrino. It should be noted, however, that both 
this non-invariance of $P_{ee}$ and large T-violating effects can only 
be observable in the (experimentally favored) 2+2 four-neutrino schemes,
in which there are two pairs of nearly degenerate neutrino mass eigenstates 
separated by a large mass gap. In the 3+1 scheme with a lone neutrino mass 
eigenstate being predominantly a sterile neutrino these effects are
expected to be small since this scheme can be considered as a small
perturbation of the three-neutrino ones.

\section*{Acknowledgements}
The authors are grateful to M. Freund, H. Igel, P. Lipari,
M. Lusignoli, and A.Yu. Smirnov for useful discussions. 
This work was supported by the ``Sonderforschungsbereich 375 f{\"u}r
Astro-Teilchenphysik der Deutschen Forschungsgemeinschaft'' (E.A, P.H.,
and M.L.) and by the Swedish Foundation for International Cooperation in
Research and Higher Education (STINT) and the Wenner-Gren Foundations
(T.O.).

\appendix
\renewcommand{\theequation}{\thesection\arabic{equation}}

\appsection
\section*{Appendix \Alph{section}: 
Analytic description of $\mbox{\boldmath$\Delta P_{ab}^T$}$ }
\label{app:2}

The evolution equation describing neutrino oscillations in the
three-flavour neutrino system can be written as
\be
i\frac{d}{dt}
\left(\begin{array}{ccc} \nu_e \\ \nu_\mu \\
\nu_\tau \end{array}\right)
=\left[U \left( \begin{array}{ccc} 0 & 0 &
0 \\ 0 & \delta & 0 \\ 0 & 0 & \Delta \end{array} \right) U^\dagger
+ \left( \begin{array}{ccc} V(t) & 0 & 0 \\ 0 & 0 & 0 \\ 0 & 0 & 0
\end{array} \right) \right] 
\left(\begin{array}{ccc} \nu_e \\ \nu_\mu \\ \nu_\tau \end{array}\right)\,,
\label{Sc1}
\ee
where $\nu_a$ ($a = e,\mu,\tau$) are the components of the neutrino
vector of state in
flavour basis, $U$ is the leptonic mixing matrix,
$\delta$ and $\Delta$ were defined in \eq~(\ref{deltas}), and
\be
V(t)=\sqrt{2}G_F N_e(t)
\label{V}
\ee
is the charged-current contribution to the matter-induced potential of
electron neutrinos, $G_F$ and $N_e$ being the Fermi weak coupling constant and
the electron number density of the medium, respectively. The
neutral-current contributions
to the potentials of $\nu_e$, $\nu_\mu$, and $\nu_\tau$ in matter are 
the same (up to tiny radiative corrections \cite{BLM}) and 
therefore do not affect the neutrino oscillations probabilities. 
The evolution matrix $S(t,t_0)$ satisfies the same
Schr\"odinger equation~(\ref{Sc1}). 

In the standard parameterization, the leptonic mixing matrix $U$ can be
written in the form\footnote{Omitting possible Majorana phases which
have no effect on neutrino oscillations.} $U=O_{23} V_{13} O_{12}$, where 
\be
O_{12} = \left( \begin{array}{ccc} c_{12} & s_{12} & 0 \\ -s_{12} & c_{12}
& 0 \\ 0 & 0 & 1 \end{array} \right),~~
V_{13} = \left(
\begin{array}{ccc} c_{13} & 0 & s_{13} e^{-i\delta_{CP}} \\ 0 & 1 & 0 \\
-s_{13} e^{i\delta_{CP}} & 0 & c_{13} \end{array} \right),~~ 
O_{23} = \left( \begin{array}{ccc} 1 & 0 & 0 \\ 0 & c_{23} & s_{23} \\
0 & -s_{23} & c_{23} \end{array} \right). 
\label{stand}
\ee
Since the matrix of matter-induced neutrino potentials ${\rm diag
\,}(V(t), 0, 0)$ 
commutes with $O_{23}$, it is convenient to perform the rotation according
to $\nu'=O_{23}^T\nu$ (or $S'=O_{23}^T S O_{23}$), where $\nu=(\nu_e~ 
\nu_\mu~\nu_\tau)^T$ and $S$ are the neutrino vector of state and evolution 
matrix in flavour basis. The effective Hamiltonian $H'(t)$ governing the
neutrino evolution in the rotated basis can be written as
\be
H'(t) = \left( \begin{array}{ccc} c_{13}^2 s_{12}^2 \delta + s_{13}^2
\Delta + V(t) & c_{13} c_{12} s_{12} \delta & c_{13} s_{13} \left(
\Delta - s_{12}^2 \delta \right) e^{-i \delta_{CP}} \\ c_{13} c_{12} s_{12}
\delta & c_{12}^2 \delta & - s_{13} c_{12} s_{12}
e^{-i \delta_{CP}} \delta \\ c_{13} s_{13} \left( \Delta - s_{12}^2 \delta
\right) e^{i \delta_{CP}} & - s_{13} c_{12} s_{12} e^{i
\delta_{CP}} \delta & s_{13}^2 s_{12}^2 \delta + c_{13}^2 \Delta \end{array}
\right).
\ee
It can be decomposed as 
\be
H'(t) = H_0'(t) + H_I'\,, \qquad H_I' = H_1' + H_2'\,,
\label{eq:hh0h1}
\ee
where
\be
H_0'(t) = 
\left(\begin{array}{ccc}
\frac{1}{2}[-\cos 2\theta_{12}\,\delta+V(t)] & \frac{1}{2}\sin
2\theta_{12}\,\delta & 0\\
\frac{1}{2}\sin 2\theta_{12}\,\delta & \frac{1}{2}[\cos
2\theta_{12}\,\delta-V(t)] & 0 \\
0 & 0 & \tilde{\Delta}(t) 
\end{array}\right)
\equiv \left( \begin{array}{cc} h(t)
& \begin{array}{c} 0 \\ 0 \end{array} \\ \begin{array}{cc} 0 & 0
\end{array} & \tilde{\Delta}(t)\end{array} \right) 
\label{eq:h0t} 
\ee
is of zeroth order in the small parameter $s_{13}$, $H_1'$ is of the first 
order, and $H_2'$ includes terms of the second and higher orders, namely
\be
H_1' = \left( \begin{array}{ccc} 0 & 0 & s_{13} \left( \Delta -
s_{12}^2 \delta \right) e^{-i \delta_{CP}} \\ 0 & 0 & - s_{13}
c_{12} s_{12} e^{-i \delta_{CP}} \delta  \\ s_{13} \left( \Delta
- s_{12}^2 \delta \right) e^{i \delta_{CP}} & - s_{13} c_{12}
s_{12} e^{i \delta_{CP}} \delta & 0 \end{array} \right)
\equiv \left( \begin{array}{ccc} 0 & 0 & a \\ 0 & 0 & b \\
a^\ast & b^\ast & 0 \end{array} \right)\, 
\label{eq:H1} 
\ee
and $H_2' = {\cal O}(s_{13}^2)$. The function $\tilde{\Delta}(t)$ that
enters into \eq~(\ref{eq:h0t}) was defined in \eq~(\ref{deltas}). We have
subtracted from the Hamiltonian a term proportional to the unit matrix
in order
to make the upper-left $2\times 2$ submatrix $h$ of $H_0'$
traceless. This amounts to multiplying all the components of the
neutrino state by the same phase factor, which does not affect the
neutrino oscillation probabilities.

We will look now for the solution to the Schr\"odinger equation for the
evolution matrix $S'$ in the rotated basis in the form 
\be
S'(t,t_0)= S_0'(t,t_0)\,S_1'(t,t_0)\,, 
\label{SS1}
\ee
where $S_0'(t,t_0)$ satisfies the equation 
\be
i\frac{d}{dt}S_0'(t,t_0)=H_0'(t) S_0'(t,t_0)\,,
\label{SS2}
\ee
with the initial condition $S_0'(t_0,t_0)=\bf{1}$. Its general solution
can be written as  
\be
S_0'(t,t_0)=\left(\begin{array}{ccc}
\alpha(t,t_0)   &  \beta(t,t_0)    & 0 \\
-\beta^*(t,t_0)  &  \alpha^*(t,t_0) & 0 \\
0 & 0 & f(t,t_0)
\end{array}\right)\,, \qquad |\alpha|^2+|\beta|^2=1\,,
\label{S0n}
\ee
where $f(t,t_0)$ is given in \eq~(\ref{f}) 
and the parameters $\alpha(t,t_0)$ and $\beta(t,t_0)$ are to be found 
from the solution of the two-flavour neutrino problem that corresponds to 
the $2\times 2$ submatrix $h$ of $H_0'$. 
%
As follows from the evolution 
equation for $S'$ and \eqs~(\ref{SS1}) and (\ref{SS2}), the matrix $S_1'$
satisfies the equation 
\be
i\frac{d}{dt}S_1'(t,t_0) = [S_0'(t,t_0)^{-1} H_I' S_0'(t,t_0)]S_1'(t,t_0)\,,
\qquad S_1'(t_0,t_0)=\bf{1}\,.
\label{S1n}
\ee 
Up to now everything is exact; we will determine now the evolution matrix 
$S'$ to the first order in perturbation theory in the small parameter 
$s_{13}$. This leads to 
\be
S'(t,t_0) \simeq S_0'(t,t_0)-iS_0'(t,t_0)\int_{t_0}^t
[S_0'(t',t_0)^{-1} H_1' S_0'(t',t_0)]\,dt'\,.
\label{Sprime}
\ee
A straightforward calculation then gives  
\be
S'(t,t_0)=\left(\begin{array}{ccc}
\alpha(t,t_0)   &  \beta(t,t_0)    & -iA \\
-\beta^*(t,t_0)  & \alpha^*(t,t_0) & -iB \\
-iC & -iD & f(t,t_0)
\end{array}\right)\,, 
\label{Sprime1}
\ee
where
\begin{eqnarray}
&& A \equiv a A_a+b A_b\,,\qquad B \equiv a B_a+b B_b\,, \\
&& C \equiv a^* C_a+b^* C_b\,,\qquad D \equiv a^* D_a+b^* D_b\,,
\label{ABCD}
\end{eqnarray}
the parameters $a$ and $b$ were defined in \eq~(\ref{eq:H1}) and 
\begin{eqnarray}
A_a &\equiv& \alpha(t,t_0)\int_{t_0}^t\alpha(t',t_0)^*f(t',t_0)\,dt'+
\beta(t,t_0)\int_{t_0}^t\beta(t',t_0)^*f(t',t_0)\,dt'\,, 
\label{Aa1} \\
A_b &\equiv& \beta(t,t_0)\int_{t_0}^t\alpha(t',t_0)f(t',t_0)\,dt'-
\alpha(t,t_0)\int_{t_0}^t\beta(t',t_0)f(t',t_0)\,dt'\,,
\label{Ab} \\
B_a &\equiv& \alpha(t,t_0)^*\int_{t_0}^t\beta(t',t_0)^*f(t',t_0)\,dt'-
\beta(t,t_0)^*\int_{t_0}^t\alpha(t',t_0)^*f(t',t_0)\,dt'\,,
\label{Ba} \\
B_b &\equiv& \alpha(t,t_0)^*\int_{t_0}^t\alpha(t',t_0)f(t',t_0)\,dt'+
\beta(t,t_0)^*\int_{t_0}^t\beta(t',t_0)f(t',t_0)\,dt'\,, 
\label{Bb} 
\end{eqnarray}
as well as 
\begin{eqnarray}
C_a &\equiv& f(t,t_0)\int_{t_0}^t\alpha(t',t_0) f(t',t_0)^*\,dt'\,,~~
C_b \equiv -f(t,t_0)\int_{t_0}^t\beta(t',t_0)^* f(t',t_0)^*\,dt', 
\label{Cab} \\ 
D_a &\equiv& f(t,t_0)\int_{t_0}^t\beta(t',t_0)f(t',t_0)^*\,dt'\,, ~~
D_b \equiv f(t,t_0)\int_{t_0}^t\alpha(t',t_0)^* f(t',t_0)^*\,dt'. 
\label{Dab} 
\end{eqnarray}
The calculation of the integrals in \eqs~(\ref{Aa1})-(\ref{Bb}) can be
simplified considerably by noticing that their right-hand sides contain the 
expressions that are the products of the elements of the evolution matrix
$S_0'(t_1,t)$. Indeed, one has 
\be 
S_0'(t_1,t)=S_0'(t_1,t_0) S_0'(t,t_0)^\dagger = 
\left(\begin{array}{ccc} 
\alpha_1 & \beta_1 & 0 \\ 
-\beta_1^* & \alpha_1^* & 0 \\ 
0 & 0 & f_1 \end{array}\right)\left(\begin{array}{ccc} 
\alpha^* & -\beta & 0 \\ 
\beta^* & \alpha & 0 \\ 
0 & 0 & f^* \end{array}\right)\,, 
\label{ss} 
\ee 
where we used the notation $\alpha\equiv\alpha(t,t_0)$, 
$\alpha_1\equiv\alpha(t_1,t_0)$, \etc~ On the other hand, we have 
\be 
S_0'(t_1,t)=\left(\begin{array}{ccc} 
\alpha(t_1,t) & \beta(t_1,t) & 0 \\ 
-\beta(t_1,t)^* & \alpha(t_1,t)^* & 0 \\ 
0 & 0 & f(t_1,t) \end{array}\right)\,.
\label{sss}
\ee
Comparing \eqs~(\ref{ss}) and (\ref{sss}) one finds 
\begin{eqnarray}
&& (\alpha\alpha_1^*+\beta\beta_1^*)f_1 f^*=\alpha(t_1,t)^*f(t_1,t)\,,
\qquad
(\beta\alpha_1-\alpha\beta_1)f_1 f^*=-\beta(t_1,t)f(t_1,t)\,, \qquad 
\label{1-2} \\
&& (\alpha^*\beta_1^*-\beta^*\alpha_1^*)f_1 f^*=\beta(t_1,t)^*f(t_1,t)\,,
\quad 
(\alpha^*\alpha_1+\beta^*\beta_1)f_1 f^*=\alpha(t_1,t)f(t_1,t)\,.
\label{3-4}
\end{eqnarray}
This allows a simplification of the integrals in \eqs~(\ref{Aa1})-(\ref{Bb}), 
making them similar in form to those in \eqs~(\ref{Cab}) and (\ref{Dab}): 
\begin{eqnarray}
&& A_a = f(t,t_0)\int_{t_0}^t\alpha(t',t)^*f(t',t)\,dt'\,, \qquad
A_b = -f(t,t_0)\int_{t_0}^t\beta(t',t)f(t',t)\,dt'\,, 
\label{Aabnew} \\
&& B_a = -f(t,t_0)\int_{t_0}^t\beta(t',t)^*f(t',t)\,dt'\,, \qquad
B_b = f(t,t_0)\int_{t_0}^t\alpha(t',t)f(t',t)\,dt'\,. 
\label{Bbnew} 
\end{eqnarray}

Rotating $S'(t,t_0)$ by $O_{23}$ back to the original flavour basis, one 
finds from \eq~(\ref{Sprime1}) 
\be
S(t,t_0)=\left(\begin{array}{ccc}
\alpha   &  c_{23}\beta-is_{23} A   & -s_{23}\beta-ic_{23} A \\
-c_{23}\beta^*-is_{23} C & S_{22} &   S_{23} \\
s_{23}\beta^*-ic_{23} C & S_{32} & S_{33} 
\end{array}\right)\,,
\label{Sfin}
\ee
where 
\begin{eqnarray}
S_{22} &\equiv& c_{23}^2\alpha^*+s_{23}^2 f-is_{23}c_{23}(B+D)\,,
\label{S22} \\
S_{23} &\equiv& -s_{23}c_{23}(\alpha^*-f)-i(c_{23}^2 B-s_{23}^2 D)\,,
\label{S23} \\
S_{32} &\equiv& -s_{23}c_{23}(\alpha^*-f)+i(s_{23}^2 B-c_{23}^2 D)\,,
\label{S32} \\
S_{33} &\equiv& s_{23}^2\alpha^*+c_{23}^2 f+is_{23}c_{23}(B+D)\,.
\label{S33}
\end{eqnarray}
The parameters $\alpha$, $\beta$, and $f$ in \eqs~(\ref{Sprime1}) and
(\ref{Sfin})-(\ref{S33}) are of zeroth order in $s_{13}$, whereas $A$, $B$,
$C$, and $D$ are of the first order. {}From \eq~(\ref{Sprime1}) one can
see that, if the mixing angle $\theta_{23}$ were zero, the T-odd probability 
difference $\Delta P^T_{\tau e}=|A|^2-|C|^2$ would have scaled with 
$\theta_{13}$ as $s_{13}^2$ 
\footnote{Due to \eq~(\ref{Tasym}) so would $\Delta P^T_{e\mu}$ and 
$\Delta P^T_{\mu\tau}$ do. However, the former cannot be found directly
from \eq~(\ref{Sprime1}) as this would require the calculation to be done in 
the next order in perturbation theory.}.  
We have checked this by solving \eq~(\ref{Sc1}) numerically. 
Since in vacuum there is no CP and T violation when any of the mixing
angles is equal to zero, non-vanishing $\Delta P_{ab}^T=|A|^2-|C|^2$ 
in the case $\theta_{23}=0$ is a pure matter effect. Moreover, it is an 
effect of asymmetric matter: as we show below, in the case of matter with
a symmetric density profile (and so also in vacuum) $|A|^2=|C|^2$,  
and all $\Delta P_{ab}^T$ vanish.
The atmospheric neutrino data indicate that $\theta_{23}$ is close 
to $45^\circ$, and therefore the T-odd probability differences must 
scale linearly with $s_{13}$. From \eq~(\ref{Sfin}) one finds
\be
\Delta P_{e\mu}^T=|S_{21}|^2-|S_{12}|^2=-2s_{23}c_{23} {\rm Im \,}
[\beta^*(A-C^*)]\,.
\label{result}
\ee
{}From the definition of the parameters $a$ and $b$ in \eq~(\ref{eq:H1})  
it follows that the ratio $|b/a|\simeq \Delta m_{21}^2/\Delta m_{31}^2$ is
small; it can also be shown that $|A_b/A_a|\sim |C_b/C_a|\sim 
\Delta m_{21}^2/\Delta m_{31}^2$. Therefore, the contributions of $A_b$ and 
$C_b$ to \eq~(\ref{result}) are suppressed by the factor $(\Delta m_{21}^2/ 
\Delta m_{31}^2)^2$, and one finally arrives at \eq~(\ref{DeltaP}). 

We will now show that our expression (\ref{result}) satisfies the
requirement that in the absence of the fundamental CP and T violation 
(\ie, in the case $\delta_{CP}=0$), the T-odd probability differences  
must vanish for any symmetric matter density profile. To do so we will show 
that in this case the expression $\beta^*(A-C^*)$ is real. 

Consider the evolution of the neutrino system over the symmetric time interval
$[-t, t]$. First, from the evolution equations (\ref{Sch3}) and
(\ref{Sch4}) we notice that in the case of symmetric matter the matrix
$S_0'(0,-t)^T$ coincides with $S_0'(t,0)$ (the phase $\delta_{CP}$ does not 
affect the evolution equations for $S_0'$). Together with \eq~(\ref{S5}) 
this leads to the following symmetry properties of $\alpha$, $\beta$,
and $f$ in symmetric matter:  
\be
\alpha(-t,0)=\alpha(t,0)^*\,,\qquad
\beta(-t,0)=\beta(t,0)^*\,,\qquad
f(-t,0)=f(t,0)^*\,.
\label{SP1}
\ee
Consider now the evolution matrix $S'(t,-t)=S'(t,0)S'(0,-t)$. It can be
parameterized in the form similar to that of \eq~(\ref{Sprime1}), and its 
entries can be found in terms of those of $S'(t,0)$ and $S'(0,-t)$. 
Using \eq~(\ref{SP1}) it is then straightforward to show that in symmetric
matter the entries of $S(t,-t)$ satisfy 
\be
A_a=C_a\,,\qquad A_b=C_b\,.
\label{SP2}
\ee
In the case of $\delta_{CP}=0$ the parameters $a$ and $b$ are real, and 
from Eqs. (\ref{ABCD}) and (\ref{SP2}) one finds $A=C$. This means that 
$A-C^*$ is pure imaginary; since $\beta$ is also pure imaginary in this
case, the right-hand side of \eq~(\ref{result}) vanishes, which completes
the proof.  

As follows from the definition of the parameters $a$ and $b$ in
\eq~(\ref{eq:H1}), they can be written as $a=e^{-i\delta_{CP}}a'$, 
$b=e^{-i\delta_{CP}}b'$ with real $a'$ and $b'$. Therefore in the general 
case of $\delta_{CP}\ne 0$ one finds that in symmetric matter $|A|=|C|$.  

\appsection
\section*{Appendix \Alph{section}: 
Two special cases}
\label{app:3}

We shall now calculate $\Delta P_{e\mu}^T$ in two special cases -- matter 
consisting of two layers of constant densities and the adiabatic 
approximation in the case of an arbitrary matter density profile. 

Consider first matter consisting of two layers of constant electron number
densities $N_1$ and $N_2$ and widths $L_1$ and $L_2$, respectively.  
The corresponding matter-induced potentials (\ref{V}) are $V_1$ and $V_2$. 
The values $\theta_1$ and $\theta_2$ of the mixing angle in the 
(1,2)-subsector in matter of densities $N_1$ and $N_2$ are given
by 
\be
\cos 2\theta_1=\frac{\cos 2\theta_{12}\delta-V_1}{2\omega_1}\,,
\qquad
\cos 2\theta_2=\frac{\cos 2\theta_{12}\delta-V_2}{2\omega_2}\,,
\label{theta1theta2}
\ee
where $\omega_1$ and $\omega_2$ were defined in \eq~(\ref{omega}). 
The time interval of the evolution of the neutrino system can be
divided into two parts: (I) $0\le t'< L_1$ and  (II) $L_1\le t'\le
L_1+L_2 \equiv L$. The parameters $\alpha$, $\beta$, and $f$ for the
first interval are given by the well-known evolution in matter of
constant density:
\begin{eqnarray}
&& \alpha(t',0)=\cos(\omega_1 t')+i\cos 2\theta_1 \sin(\omega_1 t')\,, 
\qquad 
\beta(t',0)=-i\sin 2\theta_1 \sin(\omega_1 t')\,, \\
&& 
f(t',0)=\exp(-i\Delta_1t')\,,\qquad 
~~~~~~~~~~~~~~~~~~~~~\Delta_{1,2}\equiv 
\Delta- \frac{1}{2} (\delta+V_{1,2})\,, 
\label{I}
\end{eqnarray}
whereas for the second interval they can be expressed through the 
elements of the two-flavour neutrino evolution matrix in the two-layer 
medium \cite{Akh2}:
\begin{eqnarray}
&& \alpha(t',0)=c_1c_2'-s_1 s_2'\cos(2\theta_1-2\theta_2)+i(s_1 c_2'\cos 
2\theta_1+s_2' c_1\cos 2\theta_2)\,, \\ 
&& \beta(t',0)=-i(s_1 c_2'\sin 2\theta_1+s_2' c_1\sin 2\theta_2)+
s_1 s_2'\sin(2\theta_1-2\theta_2)\,, \\
&& f(t',0)=\exp\{-i[\Delta_1 L_1+\Delta_2(t'-L_1)]\}\,.   
\label{II}
\end{eqnarray}
Here 
\be
s_2'=\sin(\omega_2\tau)\,,\qquad c_2'=\cos(\omega_2\tau)\,,\qquad
\tau=t'-L_1\,,
\label{primes}
\ee
and $s_{1,2}$ and $c_{1,2}$ were defined in \eq~(\ref{sc}). 
Direct calculation using \eqs~(\ref{DeltaP}), (\ref{Aabnew}), and (\ref{AaCa}) 
[or \eqs~(\ref{DeltaP}), (\ref{Aa1}), and (\ref{AaCa})] gives the result 
presented in \eqs~(\ref{DeltaPP})-(\ref{ZI}). 

Next, we consider the adiabatic regime in the case of an arbitrary matter
density profile. In this regime, the change of the matter density along the 
neutrino path is slow compared to the oscillation frequency. The effective
Hamiltonian of the neutrino system is approximately diagonal in the basis of 
instantaneous matter eigenstates, and the evolution of the eigenstates amounts 
to a mere multiplication by phase factors. The evolution in the flavour basis 
is obtained by rotating the evolution matrix in the instantaneous eigenstates 
basis $S(t,t_0)_{\rm eigen}$ by the matrices $U(t_0)$ and $U(t)$ of leptonic 
mixing in matter, which correspond to the initial and final times of 
neutrino evolution: $S(t,t_0)=U(t) S(t,t_0)_{\rm eigen} U(t_0)^\dag$. 
Applying this procedure to the two-flavour neutrino evolution described by 
the Hamiltonian $H_0'$, one arrives at \eqs~(\ref{alphabeta2}) and
(\ref{alphabeta}), while the parameter $f$ is always given by \eq~(\ref{f}).

We will be assuming that $V(t) \lesssim \delta\ll \Delta$, and also that
the oscillations governed by the large $\Delta=\Delta m_{31}^2/2E$ are in
the averaging regime (\ie, that $\Delta\cdot L\gg 1$). The latter
assumption allows us to obtain simple approximate expressions for the 
relevant integrals. We first note that 
\be
\int_{t_0}^t g(t') e^{-i\Delta t'}\,dt'\simeq \frac{i}{\Delta}
[g(t)e^{-i\Delta t}-g(t_0)e^{-i\Delta t_0}]+{\cal O}(1/\Delta^2)\,,
\label{approx}
\ee
where $g(x)$ is an arbitrary regular function which changes slowly on the 
time intervals $\sim 1/\Delta$, and the integration by parts has been 
used. Direct calculation then yields
\be
A_a-C_a^*\simeq
i\frac{2}{\Delta}\{\cos[\Delta(t-t_0)]-\cos(\theta-\theta_0)\cos\Phi\}\,.
\label{adiab}
\ee
\eqs~(\ref{DeltaP}), (\ref{alphabeta2}), (\ref{alphabeta}), and
(\ref{adiab}) lead to \eq~(\ref{DP2}).



\end{document}